\documentclass[aps,prd,amsmath,amssymb,showpacs]{revtex4}

\usepackage{graphicx}
\usepackage{epsfig}

\usepackage{dcolumn}
\usepackage{morefloats}
\usepackage{color}
\usepackage{slashed}
\usepackage{bbm}
\usepackage{bm}
\usepackage{multirow}

\usepackage{subfigure}
\usepackage{appendix}

\begin{document}

\title{ Hadronic loop effects on the radiative decays of the first radial excitations of $\eta$ and $\eta'$ }

\author{Yin Cheng$^{1,2}$\footnote{Email: chengyin@ihep.ac.cn}, Qiang Zhao$^{1,2}$~\footnote{E-mail: zhaoq@ihep.ac.cn} }

\affiliation{ 1) Institute of High Energy Physics,
        Chinese Academy of Sciences, Beijing 100049, P.R. China}

\affiliation{ 2) University of Chinese Academy of Sciences, Beijing 100049, P.R. China}

\begin{abstract}
Based on the one state assumption of $\eta(1405)$ and $\eta(1475)$, thus, $\eta(1295)$ and $\eta(1405/1475)$ are organized as the first radial excitations of $\eta$ and $\eta'$, respectively, we investigate the productions and radiative decays of these two states in $J/\psi\to \gamma\eta_X\to \gamma\gamma V$, where $\eta_X$ stands for $\eta(1295)$ and $\eta(1405/1475)$ and $V$ for vector mesons $\rho^0, \ \omega, \ \phi$. As we have learned from previous studies that the hadronic decays of these two states receive important contributions from the intermediate $\bar{K}K^*+c.c.$ meson loops due to the triangle singularity mechanism, we show that some measurable effects can also arise from the $\bar{K}K^*+c.c.$ meson loops in their radiative decays. Our calculation shows that the impact of the $\bar{K}K^*+c.c.$ meson loops on the $\eta(1405/1475)$ radiative decays is relatively smaller than on $\eta(1295)$ since the latter has a much larger coupling to $\bar{K}K^*+c.c.$ However, the production of $\eta(1295)$ in the $J/\psi$ radiative decays will be strongly suppressed. As a consequence of the $\bar{K}K^*+c.c.$ meson loop contributions, we find that the mixing angle extracted in the radiative decays of $\eta(1295)$ and $\eta(1405/1475)$ will be different from each other, and both are different from the one determined in other processes. 
\end{abstract}

\maketitle
\section{Introduction}
  
The study of flavor singlet and octet mixing in the lightest pseudoscalar nonet, i.e. between $\eta$ and $\eta'$, has attracted a lot of attention in the history. 
As it has been well-established that the $U(1)_A$ anomaly is the driving mechanism for many interesting phenomena for these two states, it also raises interesting questions on its role for higher radial excitation states in the isoscalar pseudoscalar spectrum, in particular, the first radial excitation states. 
The present experimental data for the $J^{P(C)}=0^{-(+)}$ spectrum are still for from satisfactory. 
For the first radial excitation there are enough states to fill a nonet between $1.25\sim 1.50$ GeV, which includes $\pi(1300)$, $K(1460)$, $\eta(1295)$ and $\eta(1405)/\eta(1475)$~\cite{Zyla:2020zbs}. 
But for higher excitations, the experimental evidences are far from well-established.
  
Even for the first radial excitations one can see that the question whether there are two states, $\eta(1405)$ and $\eta(1475)$,  present in the same mass region would have strong impact on our understanding of low-energy QCD phenomena. 
Historically, the first evidence for $\eta(1405)/\eta(1475)$ was from $p\bar{p}$ annihilations at rest into ($K\bar{K}\pi$)$\pi^+\pi^-$~\cite{Baillon:1967zz}, where a pseudoscalar of $J^{PC}=0^{-+}$ was seen in the invariant mass spectrum of $K\bar{K}\pi$.
As an SU(3) partner of the lighter pseudoscalar $\eta(1295)$, it shows that its production strength in $p\bar{p}$ annihilations is much larger than $\eta(1295)$. 
This was regarded as an evidence for its unusual flavor contents in the literature (see e.g. Ref.~\cite{Klempt:2007cp} for a review).
Later, MARK III~\cite{Bai:1990hs} and DM-2~\cite{Augustin:1990ki} reported possible two-state structures around 1.44 GeV mass region with increased statistics. 
The Obelix collaboration at LEAR \cite{Nichitiu:2002cj} seemed to confirm the MARK III result and introduced two pseudoscalars, i.e. $\eta(1405)$ and $\eta(1475)$, in the description of the invariant mass spectrum of $\eta\pi\pi$~\footnote{One notices that the masses and widths extracted from these three analyses~\cite{Bai:1990hs,Augustin:1990ki,Nichitiu:2002cj} are very different.}. 
The splitting of one state $\eta(1440)$ into two states, $\eta(1405)$ and $\eta(1475)$, suggested an outnumbering of the SU(3) nonet and could be an indication of exotic hadrons beyond the conventional quark model.
In line of this possibility there was theoretical expectation from the flux tube model that the ground state pseudoscalar glueball should have a mass around 1.4 GeV~\cite{Faddeev:2003aw}. 
It made one of these two close states, $\eta(1405)$ and $\eta(1475)$, a possible candidate for the pseudoscalar glueball, and initiated a lot of efforts on understanding their structures~\cite{Donoghue:1980hw,Close:1980rv,Barnes:1981kp,Close:1987er,Amsler:2004ps,Masoni:2006rz}. 
Moreover, its mixing with the ground state $\eta$ and $\eta'$ become an interesting topic in phenomenology although the gluon contents inside $\eta$ and $\eta'$ cannot be dramatically large~\cite{Cheng:2008ss,Tsai:2011dp,Qin:2017qes}.

While the signal for the heavier one turned out to be more clear in the $K\bar{K}\pi$ channel, and the lighter one seemed to favor the $\eta\pi\pi$ channel, $\eta(1405)$ has been assigned as the pseudoscalar glueball candidate while $\eta(1475)$ was assigned as the SU(3) partner of $\eta(1295)$ (see e.g. the mini-reviews on non-$q\bar{q}$ mesons in early editions of Particle Data Group since 1990). 
However, the glueball assignment with a low mass around 1.4 GeV is not supported by the lattice QCD (LQCD) simulations which came to the playground later. 
Both quenched~\cite{Chen:2005mg,Bali:1993fb,Morningstar:1999rf,Chowdhury:2014mra} and unquenched calculations~\cite{Richards:2010ck,Sun:2017ipk} suggest that the ground state pseudoscalar glueball should have a mass around $2.4-2.6$ GeV. 
In experiment, more and more high-precision data from $J/\psi$ and $\psi(3686)$ decays at BESIII were published during the past decade. 
There is no indication that two pseudoscalar states $\eta(1405)$ and $\eta(1475)$ are needed in the description of any exclusive channel such as $\eta\pi\pi$~\cite{Ablikim:2011pu,Ablikim:2010au,Ablikim:2019wei}, $K\bar{K}\pi$~\cite{Ablikim:2013lxa}, and the isospin-violating $3\pi$ channel~\cite{BESIII:2012aa}.
However, it seems to be true that the pseudoscalars observed in different channels have slightly shifted masses.
For instance, the mass extracted in the $K\bar{K}\pi$ channel is $1452.7\pm 3.8$ MeV~\cite{Ablikim:2013lxa}, while those in $\eta\pi\pi$ and $3\pi$ are about 1405 MeV~\cite{Ablikim:2011pu,Ablikim:2010au,Ablikim:2019wei,BESIII:2012aa}.

A breakthrough of the puzzling situation was the proposal by Ref.~\cite{Wu:2011yx} in the interpretation of the abnormally large isospin-breaking effects observed by BESIII in $J/\psi\to \gamma \eta(1405/1475)\to \gamma +3\pi$~\cite{BESIII:2012aa}. 
The interference from the intermediate $K^*\bar{K}+c.c.$ rescattering via a triangle loop can contribute to the isospin breaking at leading order due to the satisfaction of the triangle singularity (TS) condition~\cite{Landau:1959fi,Cutkosky:1960sp,bonnevay:1961aa,Peierls:1961zz}. 
The TS mechanism can naturally explain the mass shift and decay patterns with only one state around 1.4 GeV~\cite{Wu:2011yx,Wu:2012pg,Aceti:2012dj}. 
Further detailed studies including the width effects were also investigated in the literature~\cite{Achasov:2015uua,Du:2019idk}.
In Ref.~\cite{Achasov:2015uua} it was claimed that the TS contribution would be suppressed by the width effects of the intermediate $K^*$. 
Therefore, the TS mechanism may not be sufficient for accounting for the large isospin-breaking effects observed by BESIII~\cite{BESIII:2012aa}.
 A comprehensive analysis in Ref.~\cite{Du:2019idk} later showed that one important transition process via the TS mechanism was overlooked by the previous analyses. 
The TS mechanism can also enhance the direct production of $a_0(980)$ in the isospin-conserving channel, and then enhance the isospin-violating channel via the $a_0(980)$-$f_0(980)$ mixing.  
 The analysis of Ref.~\cite{Du:2019idk} thus firms up the role played by the TS mechanism in the understanding of the $\eta(1405)$ and $\eta(1475)$ puzzle.

It should be mentioned that in phenomenological studies of the pseudoscalar glueball mixing with the $q\bar{q}$ states, i.e. $\eta$-$\eta'$-$G$ or $\eta$-$\eta'$-$G$-$\eta_c$ mixings, the physical mass of the pseudoscalar glueball was assigned by $\eta(1405)$. By doing so, the gluon contents introduced into $\eta$ and $\eta'$ seem to agree with the experimental observables. However, as shown by a detailed analysis of Ref.~\cite{Qin:2017qes} following the axial vector anomaly dynamics~\cite{Cheng:2008ss,Tsai:2011dp}, the gluon contents inside $\eta$ and $\eta'$ are not sensitive to the physical mass of the pseudoscalar glueball. Furthermore, with the LQCD pure gauge glueball mass as an input, the physical mass cannot get to be lighter than 1.8 GeV~\cite{Qin:2017qes}. Similar conclusion was found by Refs.~\cite{Mathieu:2009sg,Gabadadze:1997zc} in the framework of the axial vector anomaly.

Motivated by these progresses on disentangling the $\eta(1405)$ and $\eta(1475)$ puzzle, we will investigate the scenario of treating $\eta(1295)$ and $\eta(1405)$ (Hereafter, we use $\eta(1405)$ to denote all signals related to either $\eta(1405)$ or $\eta(1475)$ in the previous two-state scenario) as the first radial excitation states of $\eta$ and $\eta'$. The radiative decay of $\eta(1405)\rightarrow\gamma \rho$ and $\eta(1405)\rightarrow \gamma \phi $ have been measured by BES-II~\cite{Bai:2004qj} and BESIII Collaboration~\cite{Ablikim:2018hxj}, respectively, in the $J/\psi$ radiative decays. We will systematically study $J/\psi\to \gamma\eta_X\to \gamma\gamma V$ with $\eta_X=\eta(1295), \ \eta(1405)$, and $V=\phi, \ \rho^0, \ \omega$, and examine the role played by the intermediate $K^*\bar{K}$ meson loops. This should provide further experimental evidences for the one-state solution for $\eta(1405)$ and $\eta(1475)$, and allow a natural categorization of $\eta(1295)$ and $\eta(1405)$ as the first radial excitation states of $\eta$ and $\eta'$. 

To proceed, we first introduce the mixing between the SU(3) flavor singlet and octet, and then present the formalism for $J/\psi\to \gamma\eta_X\to \gamma\gamma V$ in the framework of the vector meson dominance (VMD) model in Sec.~\ref{sec:2}. We stress that this will allow a self-consistent calculation of both tree-level transitions and loop corrections of the radiative decays of $\eta_X\to \gamma\gamma V$. In Sec.~\ref{sec:3} we will present our numerical results for measurable branching fractions, and discuss their phenomenological consequences. A brief summary will be given in Sec.~\ref{sec:4}. In the Appendix the loop functions for each loop transition amplitude are provided.

\section{Formalism }\label{sec:2}

\subsection{Flavor singlet and octet mixing, and parametrization for the production mechanism}

As the first radial excitation of $\eta$ and $\eta'$, $\eta(1295)$ and $\eta(1405)$ can be expressed on the quark-flavor basis similar to $\eta$-$\eta'$,
 \begin{equation}
  \begin{split}
         \eta(1295)&=\cos \alpha_P n\bar{n}- \sin\alpha_P s\bar{s}, \\
         \eta(1405)&=\sin \alpha_P n\bar{n}+ \cos \alpha_P s\bar{s},
  \end{split}
 \end{equation}
where $n\bar{n}\equiv (u\bar{u}+d\bar{d})/\sqrt{2}$, and $\alpha_P\equiv \arctan\sqrt{2}+\theta_p$ with $\theta_p$ the flavor singlet and octet mixing angle. Whether the mixing angle is the same as that for the $\eta$-$\eta'$ mixing is still an open question. In Ref.~\cite{Isgur:1975ib} Isgur proposed that although the $\eta$-$\eta'$ mixing angle  deviated from the ideal mixing significantly, the higher states should restore the ideal mixing angle given that the mass difference between the flavor $n\bar{n}$ and $s\bar{s}$ could be neglected. In this study we can either leave the mixing angle to be determined by experimental data, or test the results by adopting the same mixing angle as that for $\eta$-$\eta'$. 

Taking the advantage of the antisymmetric tensor structure of the $VVP$ coupling, we can parametrize the coupling strength for $J/\psi\to \gamma (q\bar{q})_{0^{-+}}$, where $(q\bar{q})_{0^{-+}}$ stands for a light quark-antiquark pair produced in the $J/\psi$ radiative decays, as 
\begin{eqnarray}
g_0&\equiv &\langle (q\bar{q})_{0^{-+}}|\hat{H}_\gamma|J/\psi\rangle ,
\end{eqnarray}
where $\hat{H}_\gamma$ represents the corresponding potential for the production of $(q\bar{q})_{0^{-+}}$.
Based on the SU(3) flavor symmetry the coupling strengths $\tilde{g}_{\eta_X}$ can be written as
\begin{eqnarray}\label{prod-in-jpsi-decay}
\tilde{g}_{\eta(1295)}&=& g_0(\sqrt{2} \cos\alpha_P-R \sin\alpha_P) \ ,\nonumber\\
\tilde{g}_{\eta(1405)}&=& g_0(\sqrt{2} \sin \alpha_P+  R \cos\alpha_P) \ ,
\end{eqnarray}
for $\eta(1295)$ and $\eta(1405)$, respectively. In the above equation $R$ is an SU(3) flavor symmetry breaking parameter. It distinguishes the production of an $s\bar{s}$ from $u\bar{u}$ and $d\bar{d}$ and generally takes $R\simeq m_{u/d}/m_s$.

It is clear here that the production of $\eta(1295)$ would be highly suppressed in comparison with the production of $\eta(1405)$ due to the destructive interference between the $n\bar{n}$ and $s\bar{s}$  component for $\eta(1295)$ and constructive interference for $\eta(1405)$ in the radial excitation scenario if the same mixing angle as that for the $\eta$-$\eta'$ mixing is adopted. This is consistent with the current experimental observation that signals for $\eta(1405)$ are much stronger than for $\eta(1295)$~\cite{Zyla:2020zbs}. 

Under the assumption, the ratio of $\Gamma(J/\psi \to \gamma \eta(1405))$ to $\Gamma(J/\psi \to \gamma \eta(1295))$ can be expressed as,
\begin{equation}\label{prod-ratio}
  \frac{\Gamma(J/\psi\to \gamma \eta(1405))}   {\Gamma(J/\psi\to \gamma \eta(1295))}
  =\bigg(\frac{|\mathbf{p}_{\eta(1405)}|}{|\mathbf{p}_{\eta(1295)}|}\bigg)^3
  \bigg(\frac{\sqrt{2}\sin \alpha_P+R \cos \alpha_P}{\sqrt{2}\cos \alpha_P-R \sin \alpha_P}\bigg)^2 \ ,
  \end{equation}
where the partial momentum has been included in these two $P$-wave processes. In principle, experimental data for exclusive decay branching ratios will determine $g_0$ and the relation in Eq.~(\ref{prod-ratio}) will provide a test of the radial excitation picture as $\alpha_P$ and $R$ will share the same values as for $\eta$ and $\eta'$~\cite{Zhao:2006gw,Li:2007ky}. 
However, to extract the $J/\psi$ exclusive decay branching ratios one has to subtract the decay information of the intermediate pseudoscalar mesons in $J/\psi\to\gamma\eta_X\to\gamma\gamma V$. As shown in Table~\ref{table:experimentaldata}, so far the most precise data from BESIII are still combined branching ratios. It means that a better understanding of the $\eta_X$ exclusive decay into $\gamma V$ is required.

\begin{table}
  \centering
  \small
  \caption{Branching ratios of the combined decays of $J/\psi\to\gamma \eta(1405/1475)$ and $\eta(1405/1475)$ decays into final states. For $ J/\psi \to \gamma \eta(1405)\to \gamma \gamma \phi$, two solutions are provided by the BESIII analysis~\cite{Ablikim:2018hxj}. }
  \begin{tabular}{l|c}
    \hline \hline
    Channel  & Branching ratio   \\
    \hline
    $BR(J/\psi\rightarrow \gamma \eta(1405/1475)\rightarrow\gamma K\bar{K}\pi)$ & $  (2.8\pm 0.6)\times10^{-3} $~\cite{Tanabashi:2018oca}\\
    $BR(J/\psi\rightarrow \gamma \eta(1405/1475)\rightarrow\gamma \gamma \rho)$ & $(7.8\pm 2.0)\times10^{-5}$~ \cite{Tanabashi:2018oca}\\
    $BR(J/\psi\rightarrow \gamma \eta(1405/1475)\rightarrow\gamma\gamma \phi)  \quad (I) $ &  $(7.03 \pm 0.92 \pm 0.91)\times10^{-6}$~\cite{Ablikim:2018hxj} \\ 
    $BR(J/\psi\rightarrow \gamma \eta(1405/1475)\rightarrow\gamma\gamma \phi)  \quad (II)$ &  $(10.36\pm1.51\pm1.54)\time10^{-6}$~\cite{Ablikim:2018hxj}   \\
\hline \hline
  \end{tabular}
   \label{table:experimentaldata}
\end{table}

\begin{figure}
  \centering
  \subfigure[]{\includegraphics[width=1.4in]{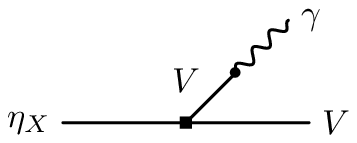}} 
  \subfigure[]{\includegraphics[width=1.5in]{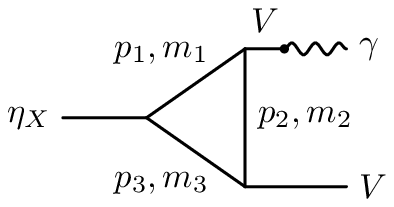}}  \\
  \subfigure[]{\includegraphics[width=1.5in]{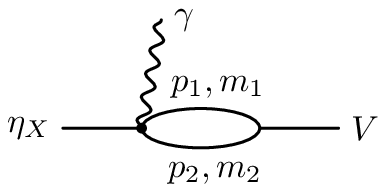}} 
  \subfigure[]{\includegraphics[width=1.5in]{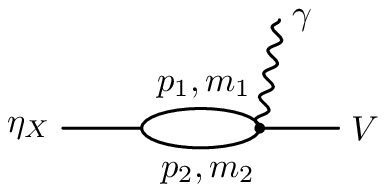}}  
  \caption{ Schematic diagrams for the process $\eta_X \to \gamma V$ at the tree and loop level in the VMD model. (a) stands for the tree-level transitions; (b) stands for the hadronic triangle loop transitions; (c) and (d) denote the contact diagrams with a photon induced by the minimum substitution at the two strong interaction vertices, respectively. }\label{fig-dynamics}
\end{figure}
        
The effective couplings for $\eta_X \to \gamma V$ are defined as: 
\begin{equation}\label{Lag-VgammaP}
         \mathcal{L}_{\eta_X\gamma V}=g_{\eta_X \gamma V} \epsilon_{\alpha \beta \delta \lambda} \partial^{\alpha}V^{\delta} \partial^{\beta}V^{\lambda}_\gamma P,
\end{equation}
where $g_{\eta_X \gamma V}$ contains the contributions from all possible mechanisms in $\eta_X \to \gamma V$. At the hadronic level the effective coupling can be decomposed into the tree diagram contributions via the VMD model, and meson loop transitions as higher order corrections. The transition mechanism is illustrated by Fig.~\ref{fig-dynamics}. Notice that the coupling vertices are well defined in the SU(3) flavor symmetry limit. The detailed calculations of the tree and loop amplitudes will be given in the next Subsection. With the calculated $g_{\eta_X \to \gamma V}$ one can express the partial decay width for $J/\psi\to \gamma \eta_X\to \gamma\gamma V$ as 
\begin{eqnarray}\label{eq:decaywidth}
  \Gamma_{J/\psi\to \gamma \eta_X\to \gamma \gamma V}  &\propto & \int \dfrac{d^3 \mathbf{p_1}d^3\mathbf{p_2}d^3\mathbf{p_3}}{(2\pi)^9 2 E_1 2 E_2 2 E_3}
 (2\pi)^4 \delta^4(P_{J/\psi}-p_1-p_2-p_3)  \nonumber\\
  &\times & \left\{ \tilde{g}_{\eta_X}    g_{\eta_X\gamma V} \left[G_{\eta_X}(s=(p_2+p_3)^2)+G_{\eta_X}(s=(p_1+p_3)^2)\right] \right\} ^2  2s \times 2 m^2_{J/\psi} |\mathbf{p_1}|^2 |\mathbf{p_2}|^2 \ ,
\end{eqnarray}
where $P_{J/\psi}$ is the four-vector momentum of the initial $J/\psi$; $G_{\eta_X}(s)$ denotes the propagator of $\eta(1405)$ or $\eta(1295)$, respectively, i.e.
\begin{equation}
  \begin{split}
  G_{\eta(1405)}(s)&=\frac{i}{s-m_{\eta(1405)}^2+i m_{\eta(1405)}\Gamma_{\eta(1405)}}, \\
  G_{\eta(1295)}(s)&=\frac{i}{s-m_{\eta(1295)}^2+i m_{\eta(1295)}\Gamma_{\eta(1295)}}.
  \end{split}
\end{equation}
Note that the parametrization of Eq.~(\ref{eq:decaywidth}) has neglected the energy or momentum dependence with the couplings $\tilde{g}_{\eta_X}$ and $g_{\eta_X\gamma V}$. For the kinematics near the pole masses of $\eta_X$, this approximation is reasonable. The decay coupling $g_{\eta_X\gamma V}$ contains contributions from both tree-level and loop amplitudes, i.e. $g_{\eta_{X}\gamma V}= g^{T}_{\eta_{X} \gamma V}+g^{L}_{\eta_{X}\gamma V}$ with the superscripts $T$ and $L$ indicating the tree and loop processes, respectively. These two quantities, $g^{T}_{\eta_{X} \gamma V}$ and $g^{L}_{\eta_{X}\gamma V}$, will then be calculated in our model.

\subsection{Tree-level amplitude in the VMD model}

In this work, we adopt the VMD model~\cite{Bauer:1977iq} to describe the electromagnetic vertices.
With the Lagrangian of Eq.~(\ref{Lag-VgammaP}) and the VMD model, the amplitude of the tree diagram shown in Fig.~\ref{fig-dynamics} (a) can be written as
\begin{eqnarray}
   i \mathcal{M}_{T}=i g^{T}_{\eta_X  \gamma V}\epsilon_{\alpha \beta \delta \lambda} p^{\alpha}_{\gamma} p^{\beta}_V \epsilon^{\delta}_{\gamma} \epsilon^{\lambda}_V,
\end{eqnarray}
where the tree-level effective coupling $g^T_{\eta_X \gamma V}$ can be expressed as:
\begin{eqnarray}
  g^{T}_{\eta_X \gamma V}   &\equiv &  -i g_{\eta_X V V} \frac{  e m^2_{V}}{f_V} G_V,  \\
\end{eqnarray}
where $G_V$ is the propagator of the intermediate vector meson $V$:
\begin{equation}
  G_V\equiv\frac{-i}{p^2_\gamma-m^2_V + i m_V \Gamma_V},
\end{equation}
with $V$ denotes the vector mesons $\rho^0, \ \omega$ or $\phi$ which is the same as the final state vector meson in $\eta_X\to \gamma V$ as required by the isospin symmetry.  The strong coupling constants $g_{\eta_X V V}$ can be extracted by other independent processes. Then, the couplings for other vectors within the same SU(3) multiplet can be related to each other by the SU(3) symmetry. We will come back to this in the next Subsection with detailed extraction of $g_{\eta_X V V}$.

The vector meson decay constant $e/f_V$ can be determined by $V\to e^+e^-$ using the experimental data, which can be expressed as,
\begin{equation}\label{vector-decay-const}
\frac{e}{f_V}=\left(\frac{3\Gamma_{V\to e^+e^-}}{2\alpha_e |{\bf p}_e|}\right)^{1/2} \ ,
\end{equation}
where ${\bf p}_e$ is the three-vector momentum of the electron in the vector-meson rest frame. The values for different vector meson decays are extracted by adopting the experimental data of the vector meson decays into $e^+e^-$~\cite{Zyla:2020zbs}, and they are listed in Table~\ref{tab-leptonicdecayconstant}.

\begin{table}
  \centering
  \caption{  Vector meson decay constants determined by $V\to e^+e^-$. The data are taken from the PDG~\cite{Zyla:2020zbs}.}\label{tab-leptonicdecayconstant}
  \begin{tabular}{c|c|c|c}
    \hline \hline
    Channel  &Total width of V & BR($V \to e^+ e^- $) & $e/f_V \ (\times 10^{-2})$ \\
   \hline
 $\phi\to e^+e^-$   &    $4.25$ MeV  &  $(2.97\pm0.04)\times 10^{-4}$    &  $-2.26$   \\
 $\rho^0\to e^+e^-$ &   $147.8$ MeV  &  $(4.72\pm 0.05 )\times 10^{-5}$  &  $6.05$   \\
 $\omega\to e^+e^-$ &    $8.49$MeV   &  $(7.36 \pm 0.15)\times 10^{-5}$  &  $1.78$  \\ 
      \hline \hline 
   \end{tabular} 
  \end{table}

\subsection{Loop amplitudes in the VMD model}

\subsubsection{Lagrangians and coupling constants}

The triangle loop amplitudes illustrated by Fig.~\ref{fig-dynamics} (b) can also be calculated in the VMD model. The loop amplitudes can reduce to an effective coupling which contributes to the $\eta_X\gamma V$ coupling in the end. Within the triangle loops the vertices for the photon couplings to the kaon and/or $K^*$ pairs can be described by the VMD model which have the following structures:
\begin{eqnarray}
 g_{K^{*+} K^- \gamma } & = & \frac{i}{\sqrt{2}} \bigg(g_{\omega K^{*+}K^-} \frac{e m^2_\omega}{f_\omega} G_\omega + g_{\rho K^{*+}K^-} \frac{e m^2_\rho}{f_\rho} G_\rho \bigg)+i g_{\phi K^{*+}K^-}\frac{e m^2_\phi}{f_\phi}  R G_\phi, \label{Kstar-to-gamma-K-charge}\\
 g_{K^{*0} \bar{K}^{0}\gamma } &=& \frac{i}{\sqrt{2}} \bigg( g_{\omega K^{*0} \bar{K}^0}\frac{e m^2_\omega}{f_\omega} G_\omega + g_{\rho K^{*0}\bar{K}^0} \frac{e m^2_\rho}{f_\rho} G_\rho \bigg)+ i g_{\phi K^{*0}\bar{K}^0} \frac{e m^2_\phi}{f_\phi} R G_\phi ,\label{Kstar-to-gamma-K-charge-neutral}
\end{eqnarray}
for the  $K^{*}K\gamma$ vertex, 
\begin{eqnarray}
  g_{K^{*+} K^{*-} \gamma } &= &\frac{i}{\sqrt{2}}\bigg( g_{\omega K^{*+}K^{*+}} \frac{ e m^2_\omega }{f_\omega}  G_\omega + g_{\rho K^{*+}K^{*+}}\frac{e m_\rho ^2}{f_\rho} G_\rho \bigg)+ i g_{\phi K^{*+} K^{*-}} \frac{e m^2_\phi }{f_\phi} R  G_\phi, \\
  g_{K^{*0}\bar{K}^{*0} \gamma } & = &\frac{i}{\sqrt{2}}\bigg(g_{\omega K^{*0}K^{*0}}\frac{ e m^2_\omega}{f_\omega} G_\omega + g_{\rho K^{*0}K^{*0}}\frac{e m^2_\rho }{f_\rho} G_\rho \bigg)+ i g_{\phi K^{*0} \bar{K}^{*0}}\frac{e m^2_\phi}{f_\phi}  R G_\phi ,
\end{eqnarray}
for the $K^*\bar{K}^* \gamma $ vertex, and
\begin{eqnarray}
  g_{K^+ K^-\gamma }  &=&  \frac{i}{\sqrt{2}} \bigg(g_{\omega K^{+}K^-}\frac{e m^2_\omega}{f_\omega} G_\omega+ g_{\rho K^{+}K^-} \frac{e m^2_\rho}{f_\rho} G_\rho \bigg)+ i g_{\phi K^{+}K^-}\frac{e m^2_\phi}{f_\phi} R G_\phi, \\
  g_{K^0 \bar{K}^0\gamma }  &=&  \frac{i}{\sqrt{2}} \bigg(g_{\omega K^0 \bar{K}^0 }\frac{e m^2_\omega}{f_\omega} G_\omega+ g_{\rho K^0\bar{K}^0} \frac{e m^2_\rho}{f_\rho} G_\rho \bigg)+ i g_{\phi K^0 \bar{K}^0}\frac{e m^2_\phi}{f_\phi} R G_\phi, \label{KKbar-to-gamma-neutral}
\end{eqnarray}
for the $K\bar{K}\gamma$ vertex. In the above equations the ground state vector meson decay constants $e/f_V$ ($V=\phi, \ \rho^0, \ \omega$) have been given by Eq.~(\ref{vector-decay-const}) and their values listed in Table~\ref{tab-leptonicdecayconstant}. 

For the hadronic vertices one can see that they can be arranged by the SU(3) symmetry. Thus, their relative strengths and phases are fixed. There are three types of hadronic coupling vertices in the loop amplitudes, i.e. $VPP$, $VVP$ and $VVV$, for which effective Lagrangians are adopted. The corresponding effective Lagrangians are as follows:
\begin{eqnarray}
 \mathcal{L}_{VPP}&=&i g_{VPP}Tr[(P\partial_{\mu}P-\partial_{\mu}P P)V^{\mu}], \label{Lagrangianvpp}\\
 \mathcal{L}_{VVP} &=&g_{VVP}\epsilon_{\alpha \beta\mu \nu} Tr[\partial^{\alpha} V^{\mu} \partial^{\beta} V^{\nu}P],\label{Lagrangianvvp} \\
 \mathcal{L}_{VVV}&=&ig_{VVV} \bigl<(\partial_{\mu}V^f_{\nu}-\partial_{\nu}V^f_{\mu})V^{\mu}V^{\nu}\bigr>, \label{Lagrangianvvv}
\end{eqnarray}
where $V$ and $P$ stand for the vector and pseudoscalar fields for the flavor SU(3) multiplets, respectively, and have the following forms:
 \begin{equation}\label{su3-pseudo}
   P=\left(
     \begin{array}{ccc}
       \frac{\sin\alpha_P \eta'+ \cos \alpha_P\eta+\pi^0}{\sqrt{2}} & \pi^{+}& K^{+}\\
       \pi^{-} & \frac{ \sin\alpha_P \eta'+ \cos \alpha_P\eta-\pi^0}{\sqrt{2}}& K^{0} \\
       K^{-} & \bar{K^{0}}& \cos\alpha_P \eta'-\sin\alpha_P \eta \\
     \end{array}
\right),
 \end{equation}
 and
 \begin{equation}\label{su3-vector}
   V=\left(
     \begin{array}{ccc}
       \frac{\omega+\rho^{0}}{\sqrt{2}} & \rho^{+} & {K^{*}}^+ \\
       \rho^{-} &  \frac{\omega-\rho^{0}}{\sqrt{2}} & K^{*0} \\
       K^{*-}& \bar{K}^{*0} & \phi \\
     \end{array}
   \right) ,
 \end{equation}
where the ideal mixing between $\omega$ ($=(u\bar{u}+d\bar{d})/\sqrt{2}$) and $\phi$ ($=s\bar{s}$) is implied. 
 
Note that we adopt the same form as Eq.~(\ref{su3-pseudo}) for the first radial excitation states of the pseudoscalar mesons.

Considering the $\eta_X$ coupling to $K^*\bar{K}$, the effective Lagrangians have the following expressions:  
   \begin{eqnarray}\label{1405triangle1}
     \mathcal{L}_{\eta(1405)K^{*+}K^-}&=&g_{\eta(1405)K^{*+}K^-}(K^- \partial_\mu \eta(1405)-\eta(1405) \partial_\mu K^{-} )(K^{*+})^\mu \nonumber\\
     &\equiv &g_{XVP} (\frac{ \sin \alpha_P}{\sqrt{2}} R -\cos \alpha_P)(K^- \partial_\mu \eta(1405)-\eta(1405) \partial_\mu K^{-} )(K^{*+})^\mu \ ,
   \end{eqnarray}
and 
   \begin{eqnarray}\label{1295triangle1}
    \mathcal{L}_{\eta(1295)K^{*+}K^-}&=&g_{\eta(1295)K^{*+}K^-}(K^- \partial_\mu \eta(1295)-\eta(1295) \partial_\mu K^{-} )(K^{*+})^\mu \nonumber\\
    &\equiv &g_{XVP} (\frac{\cos \alpha_P}{\sqrt{2}} R + \sin\alpha_P)(K^- \partial_\mu \eta(1295)-\eta(1295) \partial_\mu K^{-} )(K^{*+})^\mu,
\end{eqnarray}
where $g_{XVP}$ is the overall coupling between a radial excitation pseudoscalar $q\bar{q}$ and $K^{*+}K^-$ and will be determined later; $R$ is the SU(3) flavor symmetry breaking factor defined earlier.

Couplings between $\eta_X$ and $K^*\bar{K}^*$ also contribute to the loop amplitudes. The corresponding Lagrangians are 
\begin{eqnarray}
      \mathcal{L}_{\eta(1405) K^{*+}K^{*-}} &=&g_{XVV} \epsilon_{\alpha \beta \mu \nu }\bigg[  \frac{\sin\alpha_P}{\sqrt{2}}R \ \partial^\alpha (K^{*+})^\mu \partial^{\beta}(K^{*-})^\nu \eta(1405)+
        \cos\alpha_P \partial^\alpha (K^{*-})^\mu \partial^{\beta}(K^{*+})^\nu \eta(1405)\bigg] \nonumber\\
                        &=&g_{XVV}\bigg(\frac{\sin\alpha_P}{\sqrt{2}} R +\cos\alpha_P \bigg)  \epsilon_{\alpha \beta \mu \nu } \partial^\alpha (K^{*+})^\mu \partial^{\beta}(K^{*-})^\nu \eta(1405),
\end{eqnarray}
for the $\eta(1405)K^{*+}K^{*-}$ coupling, and 
\begin{eqnarray}
      \mathcal{L}_{\eta(1295) K^{*+}K^{*-}} &=&g_{XVV}\bigg(\frac{\cos\alpha_P}{\sqrt{2}}  R-\sin\alpha_P \bigg)  \epsilon_{\alpha \beta \mu \nu } \partial^\alpha (K^{*+})^\mu \partial^{\beta}(K^{*-})^\nu \eta(1295),
\end{eqnarray}
for the $\eta(1295)K^{*+}K^{*-}$ coupling, respectively. The coupling $g_{XVV}$ is the overall coupling strength of the first radial excitation state of a $(q\bar{q})_{0^{-+}}$ to a vector meson pair. This quantity will be determined by the combined analysis of the data for $J/\psi\to\gamma \eta(1405/1475)\to\gamma K\bar{K}\pi$ and $J/\psi\to\gamma \eta(1405/1475)\to\gamma\gamma\rho$~\cite{Zyla:2020zbs}.

The other hadronic vertices, which involve the interactions between the ground state vector and pseudoscalar mesons, can be obtained by expanding Eqs.~(\ref{Lagrangianvpp})-(\ref{Lagrangianvvv}).

We adopt the following strategy to determine the coupling constants:
\begin{itemize}
\item In our calculation we take the same sign for the ground state coupling $g_{VPP}$ and $g_{XVP}$. They are defined as positive and real numbers, and then the signs for the other couplings will be fixed. 

\item The coupling $g_{VPP}$ between the ground state vector and pseudoscalar mesons can be determined by $\phi\to K^+K^-$. Then, the other $VPP$ couplings can be related to $g_{\phi K^+ K^-}$ by the SU(3) flavor symmetry. We note that one can also extract $g_{VPP}$ via $\rho\to \pi\pi$, and some SU(3) flavor symmetry breaking effects can be found. By adopting the coupling extracted from $\phi\to K\bar{K}$, we actually absorb some leading SU(3) flavor symmetry breaking effects into this quantity since all the vertices in the loop processes involve couplings with the strange mesons. The corresponding couplings $g_{\phi K^+K^-}$ and $g_{VPP}$ are listed in Tab.~\ref{Tab:strongcoupling-VPP}. 

\item The coupling $g_{VVP}$ in the loop between the ground state vector and pseudoscalar mesons is determined by fitting the experimental data for $V\to \gamma P$ and $\eta'\to \gamma V$ in the VMD model. For these transitions between the ground state vector and pseudoscalar mesons, we assume that the intermediate ground state vector mesons saturate the transition amplitudes. The corresponding channels and fitting results are listed in Tab.~\ref{tab-excitationV}, and the best fitting gives $g_{VVP}=8.38\pm 0.1$ GeV$^{-1}$. We then adopt $g_{VVP}=8.38$ GeV$^{-1}$ to extract other $VVP$ couplings in the loop amplitudes which are listed in Tab.~\ref{Tab:strongcoupling-VVP-VVV}. The sign is determined to be consistent with the $g_{VPP}$ following the $^3P_0$ model. 

\item The strong couplings $g_{VVV}$, $g_{VVP}$ and $g_{VPP}$ can be related to each other by matching the relation of their corresponding transition amplitudes in ELA to the same relation which can be extracted explicitly in the quark pair creation (QPC) model (i.e. $^3P_0$ model). The value of $g_{VVV}$ for the relevant channels are also listed in Tab.~\ref{Tab:strongcoupling-VVP-VVV}. We mention that the two couplings $g_{VPP}$ and $g_{VVP}$ holds a reasonable relation as that extracted in the quark model, i.e. $g_{VPP}\simeq m_V g_{VVP}/2$.

\item The physical coupling $g_{\eta(1405) K^{*}\bar{K}}$ can be extracted from experimental data by a combined analysis of the $\eta(1405)$ decays into $\to K \bar{K}\pi$, $\eta\pi\pi$ and $3\pi$~\cite{Wu:2012pg,Du:2019idk}. By adopting the branching ratio $B.R.(\eta(1405)\to K^*\bar{K}+c.c.)\simeq 50\%$ and the total width $\Gamma_{\eta(1405)}=90$ MeV, we can extract $g_{XVP}$ with $\alpha_P= 42^\circ$~\cite{Du:2019idk}. Then, both $g_{\eta(1405)K^{*+}K^-}$ and $g_{\eta(1295)K^{*+}K^-}$ can be determined. Their values are listed in Tab.~\ref{Tab:strongcouplingp}. The overall coupling $g_{XVP}$ can also be determined. The extracted values are listed in Tab.~\ref{Tab:strongcouplingp}.

\item One can determine $g_{XVV}$ by treating the tree-level transition as the dominant amplitude in $\eta(1405) \to \gamma \gamma \rho$~\footnote{In principle, one can relate $g_{XVP}$ to $g_{XVV}$ in the $^3P_0$ model, which leads to $g_{XVV}\simeq 2g_{XVP}/m_X$. However, we find that this analytic relation is broken by about 30\%. This is understandable since either $K^*$ and $\bar{K}^*$ in the final state will be off-shell and the form factor corrections would become significant.}. We will show later this is reasonable for $\eta(1405)$ since the loop contributions are much smaller than the tree-level amplitude. Then, by taking the ratio of $BR(J/\psi \to \gamma \eta(1405) \to \gamma \gamma \rho)/BR(J/\psi \to \gamma \eta(1405) \to \gamma K^*\bar{K}+c.c.)$ and with the constraint from the experimental data (see Tab.~\ref{table:experimentaldata} and Refs.~\cite{Tanabashi:2018oca,Ablikim:2018hxj}), we can extract  $g_{XVV}=8.7^{+2.2}_{-1.8}$ as listed in Tab.~\ref{Tab:strongcouplingTREE}.

\end{itemize}

\begin{table}
  \centering
  \caption{Strong couplings adopted for the $VPP$ vertices in the $K^*\bar{K}(K^{(*)})$ loops.  }
  \begin{tabular}{c|c}
\hline \hline
 $VPP$ coupling const.   &    Values     \\
\hline
      $g_{\phi K^+ K^-}$         &   $ 4.47$          \\
      $g_{\rho K^+ K^-}$         &   $-\frac{g_{\phi K^+ K^-}}{\sqrt{2}}$ \\ 
      $g_{\omega K^+ K^-}$       &   $-\frac{g_{\phi K^+ K^-}}{\sqrt{2}}$ \\
      $g_{VPP}$                  &   $4.47$                               \\
      \hline \hline
    \end{tabular} 
 \label{Tab:strongcoupling-VPP}
\end{table}

\begin{table}
  \centering
  \caption{The fitted radiative transitions between the ground state vector and pseudoscalar mesons in comparison with the experimental data in the VMD model.  The best fitting gives $g_{VVP}=8.38\pm 0.1$ GeV$^{-1}$.}
  \begin{tabular}{c|c|c}
    \hline \hline 
    Channels                                         &  Experiments (MeV)                      & Fitted values (MeV)                  \\ \hline 
   $ \eta'\to \gamma \rho  $                         &  $(6.7\pm 0.7)\times 10^{-2}$           & $(8.6 \pm 0.2) \times 10^{-2}$       \\ 
   $\rho \to \gamma \eta $                           &  $(4.5\pm0.3)\times 10^{-2}$            & $(5\pm 0.1)\times 10^{-2}$           \\
   $\rho \to \gamma \pi $                            &  $(7\pm1) \times 10^{-2}$               & $ (6.1 \pm 0.1) \times10^{-2}$       \\
   $\eta' \to \gamma \omega$                         &  $(5.8\pm0.7)\times 10^{-3}$            & $(6.4 \pm0.2) \times 10^{-3}$        \\
   $\omega \to \gamma \eta$                          &  $(3.8\pm0.4) \times 10^{-3}$           & $(5.2 \pm 0.1)\times 10^{-3}$        \\
   $\omega \to \gamma \pi$                           &  $0.71 \pm 0.03$                        & $0.72  \pm 0.02$                     \\
   $\phi \to  \gamma \eta'  $                        &  $(2.6 \pm 0.1) \times 10^{-4}$         & $(2.9 \pm 0.1) \times 10^{-4}$       \\                                  
   $\phi  \to \gamma \eta  $                         &  $(5.5 \pm 0.1) \times 10^{-2}$         & $(5.2 \pm 0.1) \times 10^{-2}$           \\                                 
   $K^{*\pm}\to \gamma K^{\pm}$                      &  $(5\pm 0.5) \times 10^{-2}$            & $(2.4\pm 0.05) \times 10^{-2}$          \\
   $K^{*0}(\bar{K}^{*0})\to\gamma K^{0}(\bar{K}^0)$  &  $0.12\pm 0.01$                         & $0.081 \pm 0.002 $         \\
   \hline \hline
  \end{tabular}
  \label{tab-excitationV}
\end{table}


\begin{table}
  \centering
  \caption{Strong couplings adopted for the vertices in the $K^*\bar{K}(K^{(*)})$ loops. }
  \begin{tabular}{c|c||c|c}
\hline \hline
  VVP   &   Values (GeV$^{-1} $)  &  VVV  &   Values   \\
\hline
 $g_{\phi K^{*+}K^-}$             & $8.38$                                    &  $g_{\phi K^{*+} K^{*-}}$        &  $4.47$  \\
 $g_{\rho K^{*+}K^-}$             & $\frac{g_{\phi K^{*+}K^-}}{\sqrt{2}}$      &  $g_{\rho K^{*+}K^{*-}} $        &  $ -\frac{g_{\phi K^{*+} K^{*-}}}{\sqrt{2}}$       \\ 
 $g_{\omega K^{*+} K^-}$          & $\frac{g_{\phi K^{*+}K^-}}{\sqrt{2}}$      &  $g_{\omega K^{*+}K^{*-}} $      &  $ -\frac{g_{\phi K^{*+} K^{*-}}}{\sqrt{2}}$       \\
 $g_{VVP}$                        & $8.38$                                    &  $g_{VVV}$                       &  $4.47$  \\
      \hline \hline
    \end{tabular} 
 \label{Tab:strongcoupling-VVP-VVV}
\end{table}

\begin{table}
  \centering
  \caption{Hadronic couplings for $\eta_X\to K^{*+}K^-$ based on $g_{\eta(1405) K^{*+} K^-}=-3.64$ determined by combined analyses of the decays of $\eta(1405)\to K^*\bar{K}+c.c.$ and $\eta\pi\pi$~\cite{Wu:2012pg,Du:2019idk}. $\alpha_P= 42^\circ$ is adopted for the mixing between $\eta(1295)$ and $\eta(1405)$. }
  \begin{tabular}{c|c}
  \hline \hline
       $\eta_X VP$  & Values      \\
      \hline
      $g_{\eta(1405) K^{*+} K^-}$      &   $ -3.64$~\cite{Wu:2012pg,Du:2019idk}     \\
      $g_{\eta(1295) K^{*+} K^-}$      &   $10.9$               \\  
      $g_{XVP} $                       &   $9.98$               \\
    \hline \hline
   \end{tabular} 
   \label{Tab:strongcouplingp}
  \end{table}

\begin{table}
    \centering
    \caption{ Strong couplings for $\eta_X  \to VV$ in the VMD model. The overall coupling $g_{XVV}=8.7^{+2.2}_{-1.8}$ is extracted by the combined analysis of $J/\psi \to \gamma \eta(1405) \to \gamma \gamma \rho$ and $J/\psi \to \gamma \eta(1405) \to \gamma K^*\bar{K}+c.c.$ We adopt the central value of $g_{XVV}=8.7$ in the numerical calculation.  }
    \begin{tabular}{c|c|c}
    \hline \hline
           Coupling const. &  Expression     & Value (GeV$^{-1}$)                                                 \\
        \hline
        $g_{\eta(1295)K^{*+} K^{*-}}$     &   $g_{XVV} (\cos\alpha_P/\sqrt{2}-\sin\alpha_P)$   &  $-2.2$        \\
        $g_{\eta(1405) K^{*+}K^{*-}}$     &   $g_{XVV} (\sin\alpha_P/\sqrt{2}+\cos \alpha_P)$  &  $9.7$     \\ 
        $g_{\eta(1295) \rho \rho}$        &   $\sqrt{2} g_{XVV}\cos\alpha_P$                   &  $9.1$  \\
        $g_{\eta(1405) \rho \rho }$       &   $\sqrt{2} g_{XVV} \sin\alpha_P$                  &  $8.2$ \\  
        $g_{\eta(1295) \omega \omega}$    &   $\sqrt{2} g_{XVV} \cos\alpha_P$                  &  $9.1$  \\
        $g_{\eta(1405) \omega \omega }$   &   $\sqrt{2} g_{XVV} \sin\alpha_P$                  &  $8.2$   \\  
        $g_{\eta(1295) \phi \phi}$        &   $-2  g_{XVV} R \sin \alpha_P$                    &  $-9.3$  \\
        $g_{\eta(1405) \phi \phi }$       &   $ 2  g_{XVV} R \cos \alpha_P $                   &  $10.3$  \\
        $g_{XVV}$  & - & $8.7^{+2.2}_{-1.8}$ \\
   \hline \hline
     \end{tabular} 
     \label{Tab:strongcouplingTREE}
    \end{table}

Combining the couplings collected in Tables~\ref{tab-leptonicdecayconstant} and \ref{Tab:strongcoupling-VVP-VVV}, we can extract the effective couplings defined in Eqs.~(\ref{Kstar-to-gamma-K-charge})-(\ref{KKbar-to-gamma-neutral}) as a test of the VMD model. Since the widths of the intermediate vector mesons are considered, the couplings extracted in the VMD are complex numbers. We also provide the magnitudes of the complex couplings in round brackets in order to compare with the quantities extracted from experiment. One can see that the VMD model has accounted for the experimental data reasonably well under the assumption of the ground state vector meson saturation.

\begin{table}
  \centering 
  \caption{The effective couplings for the $K^*K^{(*)}\gamma$ transitions. The values extracted from the VMD model are listed in the second column and their corresponding modules are presented in the round brackets.  In the last column the signs of the $g_{K^*K\gamma}$ couplings are determined in the quark model. The couplings $g_{K^* K^* \gamma }$ and $g_{K K \gamma}$ are treated as pure QED couplings. Thus, their coupling strengths will be given by the charge of the hadron. Note that $e\simeq 0.3$. }
  \begin{tabular}{c|c|c}
    \hline \hline
  Electromagnetic couplings                 &  Values in VMD (Magnitude)       &   Experimental values   \\ 
     
      \hline
      $g_{K^{*+}K^+ \gamma}$  (GeV$^{-1}$)    &  $-0.17 -i 0.05 $ $(0.18)$        &    $-0.25$  \\ 
      $g_{K^{*0}K^0 \gamma}$ (GeV$^{-1}$)    &  $0.33+0.05i  $  $(0.33)$        &    $0.38$\\
      \hline   
      $g_{K^{*+}K^{*+}\gamma }$             &  $0.25 +0.026i $     $(0.25)$           &  $e$ \\
      $g_{K^{*0}K^{*0} \gamma }$            &  $-0.01 -0.024i $   $(0.03)$           &  $0$  \\
   \hline
      $g_{K^+ K^+ \gamma }$                 &  $0.25+0.026i $     $(0.25)$            &  $e$ \\      
      $g_{K^0 K^0 \gamma }$                 &  $-0.01-0.024i $  $(0.03)$            &  $0$\\
   \hline \hline
      \end{tabular}
  \label{table:Electromagneticcoupling}
\end{table}

The relations for couplings among the flavor SU(3) multiplets will be explicitly presented in the construction of each loop amplitude.

\subsubsection{Loop amplitudes}

With the above effective Lagrangians we can write down the loop transition amplitudes for Fig.~\ref{fig-dynamics} (b)-(d). 
For the triangle loops (Fig.~\ref{fig-dynamics} (b)), we use the notation $[M1,M3,(M2)]$ to denote the intermediate interaction between particle $M1$ and $M3$ by exchanging $M2$, e.g. $[K^*,\bar{K},(K)]$ represents the loop where the intermediate $K^*$ and $\bar{K}$ scattering into the final states by exchanging $K$. 
The masses and four-vector momenta of these internal particles are denoted by ($m_1$, $m_3$, $m_2$) and ($p_1$, $p_3$, $p_2$), respectively. 
The four-vector momenta of the initial-state meson $X$, final-state photon and vector meson are labelled as $p_X$, $p_\gamma $ and $p_V$, respectively.
The polarizations of the final-state photon and vector meson are $\epsilon_\gamma$ and $\epsilon_V$, respectively.
Similarly, we adopt notation $[M1,M2]$ for the internal mesons in Fig.~\ref{fig-dynamics} (c)-(d). 

In order to cut off the ultra-violet (UV) divergence in the loop integrals, we include a commonly-adopted form factor to regularize the integrand:
\begin{equation}
  \mathcal{F}(p_i^2)=\prod_i\bigg(\frac{\Lambda_i^2-m_i^2}{\Lambda_i^2-p_i^2}\bigg) \ ,
\end{equation}
where $\Lambda_i\equiv m_i + \alpha \Lambda_{QCD}$ with $\Lambda_{QCD}=220$ MeV and $\alpha=1\sim 2$ as the cut-off parameter. 

As follows, we will write down the detailed amplitude for each loop transition with explicit phase conventions.

\begin{itemize}
\item   $[K^*,\bar{K},(K)]$  
\end{itemize}

The amplitude for the $[K^*,\bar{K},(K)]$ loop is
\begin{equation}\label{KKstarKLoop}
  i\mathcal{M}= \int \frac{d^4 p_1}{(2\pi)^4}  V_{1\sigma} D^{\sigma\mu}(K^*)V_{2\mu} V_3 D(K)D(\bar{K}) \mathcal{F}(p_i^2)
  \end{equation}
where the vertex functions have been expressed by a compact form and have the following expressions,
  \begin{eqnarray}\label{vertex-func-1}
    V_{1\sigma}&=&i g_{\eta_X K^{*}\bar{K}} (p_X+p_3)_\sigma, \nonumber\\
    V_{2\mu}&=&i g_{K^* K \gamma }\epsilon _{\alpha \beta \delta \mu}  p_\gamma^\alpha p_1^\beta \epsilon_{\gamma}^\delta ,\nonumber\\
    V_3&=& i R g_{V K \bar{K}} (p_2-p_3)_\lambda \epsilon_{V}^\lambda \ ,
\end{eqnarray}
with the SU(3) flavor symmetry breaking factor $R$ included if the interacting vector meson $V= \rho $ and $\omega$. 
In Eq.~(\ref{KKstarKLoop}) functions $D^{\sigma \mu }(K^*)$ and $D(K)$ are the propagators for $K^*$ and $K$ respectively, with four-vector momentum $p$, i.e. 
\begin{eqnarray}
   D^{\sigma \mu}(K^*)&=& \frac{-i(g^{\sigma\mu}-\frac{p^\sigma p^\mu}{p^2})}{p^2-m_{K^*}^2+i\epsilon}  \nonumber\\
   D(K)&=& \frac{i}{p^2-m_{K}^2+i\epsilon}.
 \end{eqnarray}
In the calculations, the propagators will carry the corresponding four-vector momenta required by momentum conservation.

There are four isospin channels for this type of triangle diagram, which include $[K^{*+},K^{+},(K^{-})]$, $[K^{*-},K^{-},(K^{+})]$,
$[K^{*0},K^{0},(\bar{K}^{0})]$, and $[\bar{K}^{*0},\bar{K}^{0}(K^{0})]$. The vertex coupling constants are connected by the SU(3) flavor symmetry: 
\begin{equation}
  \begin{split}
    g_{\eta_{X} K^{*+} K^-}=-g_{\eta_{X} K^{*-} K^+} & =g_{\eta_{X} K^{*0} \bar{K}^0}=-g_{\eta_{X} \bar{K}^{*0} K^0},\\
     g_{\phi K^{+} K^-}=-g_{\phi K^{-} K^+} & =g_{\phi K^{0} \bar{K}^0}=-g_{\phi\bar{K}^{0} K^0}, \\
     g_{\omega K^{+} K^-}=-g_{\omega K^{-} K^+} & =g_{\omega K^{0} \bar{K}^0}=-g_{\omega\bar{K}^{0} K^0}, \\
     g_{\rho K^{+} K^-}=-g_{\rho K^{-} K^+} & =-g_{\rho K^{0} \bar{K}^0}=g_{\rho\bar{K}^{0} K^0} \ .
\end{split}
\end{equation}
The coupling constants for the sum of all the isospin channels of the $[K^*,\bar{K},(K)]$ loop can be written as the following forms for the $\gamma\phi$, $\gamma\omega$, and $\gamma\rho$ channels, respectively,
\begin{eqnarray}
  2 g_{\eta_X K^{*+} K^-} g_{\phi K^{+} K^{-}}( g_{K^{*+} K^- \gamma}+ g_{K^{*0}K^0 \gamma }), \\
  2 g_{\eta_X K^{*+} K^-} g_{\omega K^{+} K^{-}}( g_{K^{*+} K^- \gamma}+ g_{K^{*0}K^0 \gamma }), \\
  2 g_{\eta_X K^{*+} K^-} g_{\rho K^{+} K^{-}}( g_{K^{*+} K^- \gamma}- g_{K^{*0}K^0 \gamma }).
\end{eqnarray}
As listed in Tab.~\ref{table:Electromagneticcoupling}, the couplings $g_{K^{*+} K^- \gamma}$ and $g_{K^{*0}K^0 \gamma }$ have a sign phase, which implies the constructive phase between the charged and neutral meson loops for $\eta_X\to \gamma\rho$, while the $\gamma\phi$ and $\gamma\omega$ channels involve a destructive phase.

The loop integral will be given in the Appendix and the contributions of each type of the loop transitions will be collected and compared with each other among different processes. 

\begin{itemize}
\item  $[K^*,\bar{K},(K^*)]$ 
\end{itemize}

Similarly, by denoting the masses and four-vector momenta of the intermediate mesons ($K^*$, $K^*$, $\bar{K}$) as ($m_1$, $m_2$, $m_3$) and ($p_1$, $p_2$, $p_3$), respectively, the loop transition amplitude can be written as 
\begin{equation}
  i \mathcal{M}= \int \frac{d^4p_1}{(2\pi^4)} V_{1\sigma} D^{\sigma \mu }(K^*)V_{2\mu\rho} D^{\rho \nu}(K^*) V_{3\nu} D(\bar{K}) \mathcal{F}({p^2_i}) \ ,
\end{equation}
where the vertex function $V_{1\sigma}$ has the same form as that in Eq.~(\ref{vertex-func-1}), and the other two functions are: 
\begin{eqnarray}
    V_{2\mu \rho} &=&-i g_{K^* K^* \gamma } \bigg[\epsilon^{\delta}_{\gamma}g_{\delta\mu}p_{1\rho} +\epsilon^\delta_\gamma g_{\delta \rho}p_{2\mu} - g_{\mu\rho}(p_1+p_2)_\delta \epsilon^\delta_\gamma \bigg], \\
    V_{3\nu}      &=& i R g_{V K^{*+}K^-} \epsilon_{\alpha\beta\nu\lambda} p_2^\alpha p_V^\beta \epsilon^\lambda_V.  
\end{eqnarray}

Similar to the previous loop amplitude, there are also four isospin channels which involve the charged and neutral intermediate mesons. They can be combined together with the coupling constants for the $\gamma \phi$, $\gamma\omega$ and $\gamma\rho^0$ channels, respectively, i.e.
 \begin{eqnarray}
  2 g_{\eta_{X} K^{*+} K^-} g_{\phi K^{*+} K^-}(g_{K^{*+} K^{*-} \gamma }+g_{K^{*0}\bar{K}^{*0} \gamma }) , \\
  2 g_{\eta_{X} K^{*+} K^-} g_{\omega K^{*+} K^-}(g_{K^{*+} K^{*-} \gamma }+g_{K^{*0}\bar{K}^{*0} \gamma }) , \\
  2 g_{\eta_{X} K^{*+} K^-} g_{\rho K^{*+} K^-}(g_{K^{*+} K^{*-} \gamma }-g_{K^{*0}\bar{K}^{*0} \gamma }) .
 \end{eqnarray}
As listed in Tab.~\ref{table:Electromagneticcoupling}, the charge neutral coupling $g_{K^{*0}\bar{K}^{*0}\gamma} $ is much  smaller than the charged one. 
   
\begin{itemize}
\item   $[K^*,\bar{K}^*,(K)]$ 
\end{itemize}

With the same notation convention for the masses and four-vector momenta of the intermediate mesons, the loop amplitude can be expressed as 
\begin{equation}
  i \mathcal{M}=\int \frac{d^4 p_1}{(2\pi)^2} V_{1\mu\nu} D^{\mu \mu'}(K*) V_{2\mu'} D(K) V_{3\nu'} D^{\nu'\nu}(\bar{K}^*) \mathcal{F}(p^2_i) \ ,
\end{equation} 
where the vertex function $V_{2\mu'}$ has been given in Eq.~(\ref{vertex-func-1}) and the other two vertex functions are
\begin{eqnarray}
V_{1\mu\nu} &=& -i g_{\eta_X K^* \bar{K}^*} \epsilon_{\alpha\beta\mu\nu} p^\alpha_1 p^\beta_3 \\
V_{3\nu'} &=& i R g_{V K^* \bar{K}}\epsilon_{\alpha_2 \beta_2 \nu' \lambda} p^{\alpha_2}_3 p^{\beta_2}_V \epsilon^{\lambda}_{V} \ .
\end{eqnarray}
Combining together the four isospin channels, we have the coupling constants for the $\gamma\phi$, $\gamma\omega$, and $\gamma\rho^0$ decays, respectively, as follows:
\begin{eqnarray}
  2 g_{\eta_X K^{*+} K^{*-}} g_{\phi K^{*+} K^{-}}( g_{K^{*+} K^- \gamma}+ g_{K^{*0}K^0 \gamma }), \\
  2 g_{\eta_X K^{*+} K^{*-}} g_{\omega K^{*+} K^{-}}( g_{K^{*+} K^- \gamma}+ g_{K^{*0}K^0 \gamma }), \\
  2 g_{\eta_X K^{*+} K^{*-}} g_{\rho K^{*+} K^{-}}( g_{K^{*+} K^- \gamma}- g_{K^{*0}K^0 \gamma }).
\end{eqnarray}

\begin{itemize}
\item  $[K^*,\bar{K}^*,(K^*)]$ 
\end{itemize}
    
The transition amplitude can be written as
\begin{equation}
  i\mathcal{M}=\int \frac{d^4 p_1}{(2\pi)^4} V_{1\mu\nu} D^{\mu \mu'}(K^*) V_{2\mu'\rho} D^{\rho \sigma}(K^*) V_{3\nu'\sigma} D^{\nu'\nu}(\bar{K}^*) \mathcal{F}(p^2_i),
\end{equation}
where the vertex functions have been given earlier for the corresponding couplings. Similar to the loop amplitudes discussed, we combine the isospin channels together to give the couplings for the $\gamma\phi$, $\gamma\omega$, and $\gamma\rho^0$ decays, respectively, as follows:
\begin{eqnarray}
  2 g_{\eta_{X} K^{*+} K^{*-}} g_{\phi K^{*+} K^{*-}}(g_{K^{*+} K^{*-} \gamma }+g_{K^{*0}\bar{K}^{*0} \gamma }) ,\\
  2 g_{\eta_{X} K^{*+} K^{*-}} g_{\omega K^{*+} K^{*-}}(g_{K^{*+} K^{*-} \gamma }+g_{K^{*0}\bar{K}^{*0} \gamma }) ,\\
  2 g_{\eta_{X} K^{*+} K^{*-}} g_{\rho K^{*+} K^{*-}}(g_{K^{*+} K^{*-} \gamma }-g_{K^{*0}\bar{K}^{*0} \gamma }) .
\end{eqnarray}

\begin{itemize}
\item  $[K,\bar{K}^*,(K)]$ 
\end{itemize}

The transition amplitude can be written as
\begin{equation}
  i \mathcal{M} = \int \frac{d^4 p_1}{(2\pi)^4} V_{1\mu} D(K) V_2 D(K) V_{3\nu} D^{\mu\nu}(\bar{K}^*) \mathcal{F}(p^2_i) \ ,
\end{equation}
where the vertex functions $V_{1\mu}$ and $V_{3\nu}$ have been given earlier, and $V_2$ has the following form:
\begin{eqnarray}
           V_2      &=&i g_{K K \gamma } (p_1+p_2)_\delta \epsilon^{\delta}_\gamma \ .
\end{eqnarray}

Again, combining the four isospin channels of this type of the loop transitions, we obtain the coupling constants for  the $\gamma\phi$, $\gamma\omega$, and $\gamma\rho^0$ decays, respectively, as follows:
\begin{eqnarray}
  2 g_{\eta_X K^{+} K^{*-}}g_{\phi K^{+} K^{*-}}(g_{K^+ K^+ \gamma}+g_{K^0 K^0 \gamma})  , \\
  2 g_{\eta_X K^{+} K^{*-}}g_{\omega K^{+} K^{*-}}(g_{K^+ K^+ \gamma}+g_{K^0 K^0 \gamma}) , \\
  2 g_{\eta_X K^{+} K^{*-}}g_{\rho K^{+} K^{*-}}(g_{K^+ K^+ \gamma}-g_{K^0 K^0 \gamma}) ,
 \end{eqnarray}
where the charge neutral amplitudes vanish literally due to the suppressed coupling $g_{K^0 K^0 \gamma}$.

\begin{itemize}
\item  $[K,\bar{K}^*,(K^*)]$ 
\end{itemize}

The transition amplitude can be written as
\begin{equation}
 i \mathcal{M}= \int \frac{d^4 p_1}{(2\pi)^4} V_{1\mu} D(K) V_{2\nu} D^{\nu \nu'}(K^*) V_{3\mu'\nu'} D^{\mu \mu'}(\bar{K}^*) \mathcal{F}(p^2_i) \ ,
\end{equation}
where all the vertex functions have been given earlier. The combined couplings for these four isospin channels can be expressed as the following forms for  the $\gamma\phi$, $\gamma\omega$, and $\gamma\rho^0$ decays, respectively, 
 \begin{eqnarray}
  2 g_{\eta_{X} K^{+} K^{*-}} g_{\phi K^{*+} K^{*-}}(g_{K^{*+}K^+\gamma}+g_{K^{*0}K^0 \gamma}) , \\
  2 g_{\eta_{X} K^{+} K^{*-}} g_{\omega K^{*+} K^{*-}}(g_{K^{*+}K^+\gamma}+g_{K^{*0}K^0 \gamma}) ,\\
  2 g_{\eta_{X} K^{+} K^{*-}} g_{\rho K^{*+} K^{*-}}(g_{K^{*+}K^+\gamma}-g_{K^{*0}K^0 \gamma}) .
 \end{eqnarray}

\begin{itemize}
\item Contact loop diagrams 
\end{itemize}

Besides the triangle loop transitions, the photon can be produced by the minimum substitution of the derivative at the hadronic interaction vertices, i.e. 
\begin{equation}
   \partial_\mu \rightarrow \partial _\mu + i e \hat{Q} A_\mu.
\end{equation}
Such contributions are illustrated by Fig.~\ref{fig-dynamics} (c) and (d) and are referred to as the contact loop diagrams in this paper.

For Fig.~\ref{fig-dynamics} (c) or (d), the intermediate mesons $[M1,M2]$ both could be $[K^*, \bar{K}]$ or $[K^*, \bar{K}^*]$. It can be easily proved that contributions from Fig.~\ref{fig-dynamics} (c) with either $[K^*, \bar{K}]$ or $[K^*, \bar{K}^*]$ should vanish~\cite{Li:2007xr}. For Fig.~\ref{fig-dynamics} (d) with $[M1,M2]=[K^*,\bar{K}^*]$, the antisymmetric tensor coupling for the $\eta_X K^* \bar{K}^*$ vertex dictates that the induced photon can only contribute via its longitudinal component. Thus, for the real photon decay, this transition will be forbidden. As a result, only the transition of Fig. \ref{fig-dynamics} (d) with the intermediate $K^*\bar{K}$ mesons will have non-vanishing contributions and the corresponding amplitude can be expressed as follows:
\begin{equation}
   i \mathcal{M}=\int \frac{d^4 p_1}{(2\pi)^4} V_{1\mu} D^{\mu \alpha}(K^{*+}) V_{2\alpha} D(K^{-}) \mathcal{F}(p^2_i),
\end{equation}
where the vertex function $V_{1\mu}$ has been defined in Eq.~(\ref{vertex-func-1}) and $V_{2\alpha}$ has the following form:
\begin{eqnarray}
        V_{2\alpha} &=&-i e (\hat{Q}K^{*+})   R  g_{V K^{*+}K^{-}} \epsilon_{\delta \beta \alpha \lambda } \epsilon^{\delta}_\gamma p_V^\beta \epsilon^{\lambda}_V.
\end{eqnarray}
It is obvious that the intermediate charge neutral loops cannot contribute since the amplitude is proportional to the charge of the pseudoscalar meson. Moreover, taking into account the SU(3) relation among the couplings, one finds that the two amplitudes with the intermediate $K^{*-}K^+$ and $K^{*+}K^-$ loops are constructive for all the decay channels into $\gamma\phi$, $\gamma\omega$ and $\gamma\rho^0$.

The detailed expressions of the loop amplitudes defined in this Subsection are provided in the Appendix. Meanwhile, taking the advantage of the antisymmetric tensor structure for the $VVP$ ($V\gamma P$) coupling, we can define effective couplings for each transition amplitude and the total amplitude can be written as a sum of all these amplitudes, i.e.
\begin{eqnarray}
  i \mathcal{M}_{Total}&=&i(g^{T}_{\eta_X \gamma V}  + g^{L}_{\eta_X \gamma V} )\epsilon_{\alpha\beta\delta \lambda} p^\alpha_{\gamma} p^{\beta}_{V}\epsilon^{\delta}_{\gamma}\epsilon_{V}^{\lambda} \ ,
\end{eqnarray} 
where $g^{T}_{\eta_X \gamma V}$ and  $g^{L}_{\eta_X \gamma V}$ are the effective couplings extracted from the tree and loop transitions, respectively. In order to understand better the behavior of the loop transitions, we introduce a switch $\delta_C$ to rewrite $g^{L}_{\eta_X \gamma V}$ as
\begin{equation}
g^{L}_{\eta_X \gamma V}\equiv g^{L}_{triangle}+\delta_C g^{L}_{contact} \ .
\end{equation}
With $\delta_C=1$ or 0, one is able to switch on/off the contributions from the contact loop diagram in the calculation.

\section{Numerical results and discussions}\label{sec:3}

The present experimental measurements of the exclusive decays of $\eta(1295)$ and $\eta(1405)$ are still far from satisfactory. In particular, the mass degeneracy of $f_1(1285)$ and $f_1(1420)$ with $\eta(1295)$ and $\eta(1405)$ has brought in a lot of challenges to the data analysis. At this moment, the data from $J/\psi$ and $\psi(3686)$ decays provide the joint branching ratios for the production and decay of these two states. 

\subsection{Partial decay widths for $\eta(1405)$ and $\eta(1295)\to \gamma V$}

With the amplitudes and parameters provided in the previous Section, we can directly calculate the radiative decays of these two states. In Tab.~\ref{pred-etaX-to-gamma} the exclusive contributions from the tree-level amplitudes are listed and compared with the results with the loop contributions included. Two values for the cut-off parameter $\alpha$ are adopted for the loop amplitudes to show the sensitivities of the partial decay widths to the loop contributions. One can see that the loop transitions have relatively small impact on the $\eta_X\to\gamma \rho^0$ decay channels, but play a significantly role in $\eta_X\to \gamma\omega$ and $\gamma\phi$. Such an interference effect can be further investigated by the branching ratio fractions between two exclusive decay channels in the next Subsection.

\begin{table}
  \centering 
  \caption{ Partial decay widths of $\eta_X \to \gamma V$ (in unit of MeV) calculated by the exclusive tree-level amplitudes, and two values $\alpha=1$ and $2$ for the cut-off parameter are adopted as a comparison. }\label{pred-etaX-to-gamma}
  \begin{tabular}{c|c|c|c|c}
    \hline \hline
  & Channels           &  Widths (tree amp.)   & Widths (all) with $\alpha=1$  & Widths (all) with $\alpha=2$      \\ \hline
    \multirow{3}*{$\eta(1295)$}             
                                            &  $\gamma \rho$    &   $1.72$      &    $1.76$              &      $1.45$               \\
                                            &  $\gamma \omega$  &   $0.15$      &    $0.23$              &      $0.38$               \\ 
                                            &  $\gamma \phi$    &   $0.05$      &    $0.09$              &      $0.16$               \\\hline

    \multirow{3}*{$\eta(1405)$}             &  $\gamma \rho$    &   $2.25$      &    $2.26$              &      $2.67$               \\ 
                                            &  $\gamma \omega$  &   $0.19$      &    $0.16$              &      $0.2$              \\ 
                                            &  $\gamma \phi$    &   $0.16$      &    $0.24$              &      $0.35$               \\\hline \hline
                                          \end{tabular}
  \end{table}  

We can also examine the exclusive contributions from each loop diagram in order to clarify their roles in the interference with the tree-level amplitude. The results are listed in Tab.~\ref{tab-loop} for two cut-off values, i.e. $\alpha=1$ and $\alpha=2$. It shows that the triangle diagrams (Fig.~\ref{fig-dynamics} (b)) have much smaller contributions than the contact diagrams (Figs.~\ref{fig-dynamics} (c) and (d)), and the decay channel $\eta(1295)\to \gamma \omega$ has the largest contributions from the loop diagrams. This will lead to significant changes to the 
the branching ratio fraction between the two channels of $\eta(1295)\to \gamma\phi$ and $\gamma\omega$. In addition, we also see some sensitivities of the loop contributions to the cut-off parameter $\alpha$. It is slightly surprising but understandable that the loop transition corrections are significant for $\eta(1295)$. Note that it has large couplings to $K^*\bar{K}+c.c.$ in the mixing scheme which will strongly enhance the loop amplitude. Moreover, the tree-level amplitude is suppressed by the intermediate $\omega\to e^+e^-$ coupling in the VMD model.

\begin{table}
  \centering
  \caption{ Decay widths of each type of the hadronic loop diagrams with the cut-off parameter $\alpha=1$ and $2$.}
  \begin{tabular}{c|c|cc|cc}
    \hline \hline
     \multirow{2}*{Diagrams}& \multirow{2}*{Decay channels (KeV)} &  \multicolumn{2}{c|}{$\alpha=1$}                        &  \multicolumn{2}{c}{$\alpha=2$}  \\  \cline{3-6}
             
                                                &                      &   $\eta(1295)$  &    $\eta(1405)$   &    $\eta(1295)$   &  $\eta(1405)$ \\ \hline
        \multirow{3}*{ $[K^*,\bar{K},(K)]$  } &    $\gamma \rho$     &      $0.37$               &          $0.21$             &           $4.09$             &           $1.64$          \\
                                                &    $\gamma \omega$   &      $0.03$               &          $0.018$            &           $0.35$             &           $0.14$         \\ 
                                                &    $\gamma \phi $    &      $0.26$               &          $0.4$              &           $1.23$             &           $1.1$          \\ \hline

      \multirow{3}*{$[K^*,\bar{K},(K^*)]$ }   &    $\gamma \rho$     &      $0.06$               &          $0.03$             &           $1.1$              &           $0.38$         \\
                                                &    $\gamma \omega$   &      $0.05$               &          $0.024$            &           $0.84$             &           $0.3$         \\ 
                                                &    $\gamma \phi $    &      $0.04$               &          $0.03$             &           $0.8$              &           $0.4$         \\ \hline

      \multirow{3}*{$[K^*,\bar{K}^*,(K)]$ }     &    $\gamma \rho$     &      $9.7\times10^{-5}$     &       $5.6 \times 10^{-3}$  &       $2.4\times10^{-3}$     &     $0.13$               \\
                                                &    $\gamma \omega$   &      $8.6\times 10^{-6}$  &       $5.0\times 10^{-4}$   &       $2.1\times10^{-4}$     &     $0.012$               \\              
                                                &    $\gamma \phi $    &      $1.5\times10^{-5}$   &       $1.3\times 10^{-3}$   &       $3.0\times10^{-4}$     &     $0.021$               \\   \hline

      \multirow{3}*{$[K^*,\bar{K}^*,(K^*)]$ } &    $\gamma \rho$     &     $0.03$                &       $1.29$                 &         $0.26$               &      $10.5$              \\
                                                &    $\gamma \omega$   &     $0.024$               &       $1.03$                  &         $0.2$                &      $8.4$              \\
                                                &    $\gamma \phi $    &     $0.02$               &       $1.37$                 &         $0.17$               &      $10.4$                \\  \hline

      \multirow{3}*{$[K,\bar{K}^*,(K)]$ }     &    $\gamma \rho$     &       $0.19$              &       $0.1$                 &         $2.6$               &       $0.99$             \\
                                                &    $\gamma \omega$   &       $0.15$              &       $0.08$                &         $2.1$               &       $0.8$             \\ 
                                                &    $\gamma \phi $    &       $0.2$              &       $0.18$                 &         $2.4$               &       $1.46$              \\ \hline
       
      \multirow{3}*{$[K,\bar{K}^*,(K^*)]$ }   &    $\gamma \rho$     &       $0.23$              &       $0.07$                 &         $5.4$                &       $1.33$             \\
                                                &    $\gamma \omega$   &       $0.02$              &       $6.2 \times 10^{-3}$   &         $0.49$               &       $0.12$             \\ 
                                                &    $\gamma \phi $    &       $0.03$              &       $0.012$                &         $0.72$                &       $0.25$             \\ \hline
            
      \multirow{3}*{$[K^*,\bar{K}]$  }         &    $\gamma \rho$     &       $25.1$              &       $6.6$                 &         $188.2$              &       $41.9$               \\
                                                &    $\gamma \omega$   &       $23.8$              &       $6.3$                 &         $178.7$              &       $40.3$               \\ 
                                                &    $\gamma \phi $    &       $15.9$              &       $6.4$                  &         $119.1$              &       $40.5$               \\ \hline 
             
       \multirow{3}*{All loops}        &    $\gamma \rho$               &       $24.9$                &       $5.5$                &        $193.8$               &        $59.5$                    \\
                                                &    $\gamma \omega$   &       $25.7$                 &        $5.0$                &        $179.4$              &        $37.3$                    \\
                                                &    $\gamma \phi$     &       $15.68$                &       $11.0$                &        $105.6$               &        $58.2$                   \\
                                                \hline \hline
                                           \end{tabular}
\label{tab-loop}
\end{table}

\subsection{Relative production rate between $\eta(1405)$ and $\eta(1295)$}

In order to extract information about their internal structures, we define several branching ratio fractions which can be directly compared with the experimental data. For the production and decay of $\eta(1295)$ and $\eta(1405)$ in the same channel, we define
\begin{eqnarray}\label{eq:definitionRatiosamef}
  \mathcal{R}_{\rho}&\equiv &\frac{\Gamma_{J/\psi\to \gamma \eta(1405)\to \gamma \gamma \rho}}{\Gamma_{J/\psi\to \gamma \eta(1295)\to \gamma \gamma \rho}}, \nonumber\\
  \mathcal{R}_{\omega}&\equiv &\frac{\Gamma_{J/\psi\to \gamma \eta(1405)\to \gamma \gamma \omega}}{\Gamma_{J/\psi\to \gamma \eta(1295)\to \gamma \gamma \omega}}, \nonumber\\ 
  \mathcal{R}_{\phi}&\equiv &\frac{\Gamma_{J/\psi\to \gamma \eta(1405)\to \gamma \gamma \phi}}{\Gamma_{J/\psi\to \gamma \eta(1295)\to \gamma \gamma \phi}} .
\end{eqnarray}

With the amplitudes given in the previous Section, we can first examine the branching ratio fractions contributed by the tree diagram in Fig.~\ref{fig-dynamics}.  
As an example, the fraction $\mathcal{R}_{\rho}$ has the following expression from the tree-level transitions:
\begin{eqnarray}
\mathcal{R}_{\rho}&=&\bigg(\frac{|\mathbf{p}_{\eta(1405)}|}{|\mathbf{p}_{\eta(1295)}|}\bigg)^3\bigg(\frac{|\mathbf{p}_{\rho}|}{|\mathbf{p}_{\rho}'|}\bigg)^3\frac{m_L\Gamma_L}{m_H\Gamma_H}\bigg(\frac{\sqrt{2}\tan\alpha_P+R}{\sqrt{2}-R\tan\alpha_P}\bigg)^2\tan^2\alpha_P \ ,
\end{eqnarray}
where $\mathbf{p}_{\eta(1405)}$ and $\mathbf{p}_{\eta(1295)}$ denote the three-vector momenta of $\eta(1405)$ and $\eta(1295)$ in the $J/\psi$ rest frame, respectively; $\mathbf{p}_{\rho}$ and $\mathbf{p}_{\rho}'$ are the final state vector meson momenta in the rest frame of $\eta(1405)$ and $\eta(1295)$, respectively. This ratio is likely to be larger than unity given that $\alpha_P\simeq 42^\circ$ in the scenario of the first radial excitations. Similarly, the ratios $\mathcal{R}_{\omega}$ and $\mathcal{R}_{\phi}$ from the tree diagrams can be extracted. 

In Tab.~\ref{tab:RrhoRphiplus1} these three branching ratio fractions from the tree diagrams are listed. It shows that the combined branching ratio for $\eta(1405)$ is about one order of magnitude larger than that for $\eta(1295)$. This is consistent with the experimental observations that the signal for $\eta(1295)$ is significantly suppressed in the $J/\psi$ radiative decays. 

\begin{table}
  \centering
 \caption{ Numerical results of $\mathcal{R}_\rho$, $\mathcal{R}_\omega$ and $\mathcal{R}_\phi$ with SU(3) breaking factor $R=0.8$ and mixing angle $\alpha_P=42^\circ$.}
 \begin{tabular}{c|ccc}
  \hline\hline
 $\alpha=1$  &  $\mathcal{R}_\rho$  & $\mathcal{R}_\omega$  &  $\mathcal{R}_\phi$   \\
  \hline
  Tree level              &       $10.8$                      &              $10.9$               &   $20.8$    \\
  T+L, $\delta_C=0$       &       $11.7$                      &              $10.6$               &   $24.0$    \\
  T+L, $\delta_C=1$       &       $9.2$                       &              $5.5$                &   $16.3$    \\
  \hline \hline   
  \end{tabular}
    \label{tab:RrhoRphiplus1}
\end{table}

The inclusion of the loop processes introduce sizeable corrections to the branching ratio fractions as we have learned earlier. In Tab.~\ref{tab:RrhoRphiplus1} the calculation results of $\mathcal{R}_\rho$, $\mathcal{R}_\omega$ and $\mathcal{R}_\phi$ with the loop contributions are also listed. The results with $\delta_C=1$ and $\delta_C=0$ distinguish the situations whether or not to include the contributions from Fig.~\ref{fig-dynamics} (d) in the loop amplitude, respectively. Note again that Fig.~\ref{fig-dynamics} (c) always vanishes. It shows that the inclusion of the contact loop diagram has led to significant corrections to both $\mathcal{R}_\omega$ and $\mathcal{R}_\phi$. In contrast, $\mathcal{R}_\rho$ appears to be a relatively  stable  quantity. This is due to the relatively large contributions from the tree diagrams in the $\gamma \rho$ channel for both $\eta(1405)$ and $\eta(1295)$. In particular, the tree diagram contributions are dominant in $\eta(1405)\to \gamma\rho^0$.

\subsection{Branching ratio fractions between different radiative decay channels}

For $\eta(1295)$ or $\eta(1405)$ radiative decays into different channels we define the following branching ratio fractions:
\begin{eqnarray}\label{eq:definitionRatiodifferf}
  \mathcal{R}^{\rho/\phi}_{\eta(1405)}&\equiv& \frac{\Gamma_{ \eta(1405)\to \gamma \rho}}{\Gamma_{\eta(1405)\to \gamma \phi}}, \nonumber\\ 
  \mathcal{R}^{\phi/\omega}_{\eta(1405)}&\equiv& \frac{\Gamma_{ \eta(1405)\to \gamma \phi}}{\Gamma_{\eta(1405)\to \gamma \omega}}, \nonumber\\
  \mathcal{R}^{\rho/\phi}_{\eta(1295)}&\equiv& \frac{\Gamma_{ \eta(1295)\to \gamma \rho}}{\Gamma_{\eta(1295)\to \gamma \phi}}, \nonumber\\ 
  \mathcal{R}^{\omega/\phi}_{\eta(1295)} &\equiv& \frac{\Gamma_{ \eta(1295)\to \gamma \omega}}{\Gamma_{\eta(1295)\to \gamma \phi}} \ .
 \end{eqnarray}  
Note that the fractions in Eq.~(\ref{eq:definitionRatiodifferf}) is defined in such a way that the values at leading order will be larger than unity if $\eta(1295)$ and $\eta(1405)$ are the first radial excitations of $\eta$ and $\eta'$, respectively. 

Supposing that only the tree-level amplitudes contribute in $\eta_X\to\gamma V$, the branching ratio fractions defined in Eq.~(\ref{eq:definitionRatiodifferf}) would have simple forms in terms of the mixing angle. As an example, the ratio between the $\gamma\rho^0$ and $\gamma\phi$ decay channel can be written as:
\begin{eqnarray}\label{eq:rate-1405-rho-phi}
  \mathcal{R}^{\rho/\phi}_{\eta(1405)}&\equiv& \frac{\Gamma_{ \eta(1405)\to \gamma \rho}}{\Gamma_{\eta(1405)\to \gamma \phi}}=\left(\frac{|\mathbf{p}_\rho|}{|\mathbf{p}_\phi|}\right)^3\left[\frac{(e m_\rho^2/f_\rho)G_\rho}{(e m_\phi^2/f_\phi)G_\phi}\right]^2\frac{\tan^2\alpha_P}{2R^2} \ , 
 \end{eqnarray}  
where $\mathbf{p}_\rho$ and $\mathbf{p}_\phi$ are the three-vector momenta of the final state $\rho^0$ and $\phi$ in the initial $\eta(1405)$ rest frame. By including the loop amplitudes, the branching ratio fractions will deviate from the above expectation.

As mentioned in the Introduction that the BESIII Collaboration recently measured the radiative decay of $\eta(1405)\to \gamma\phi$ in $J/\psi\to \gamma\gamma\phi$~\cite{Ablikim:2018hxj}, we can thus calculate $\mathcal{R}^{\rho/\phi}_{\eta(1405)}$ and compare it with the data. The results are presented in Fig.~\ref{fig-rho-phi-eta1405} in terms of two parameters $\alpha_P$ and $m_{\eta(1405)}$. These two variables are closely related to the interpretation of these two pseudoscalars. Therefore, the dependence of $\mathcal{R}^{\rho/\phi}_{\eta(1405)}$ on these two variables can illustrate whether it is a reasonable picture to treat these two states as the first radial excitation states. 

\subsubsection*{$\eta(1405)\to \gamma V$}

\begin{figure}
  \centering
    \subfigure[]{\includegraphics[width=2.5 in]{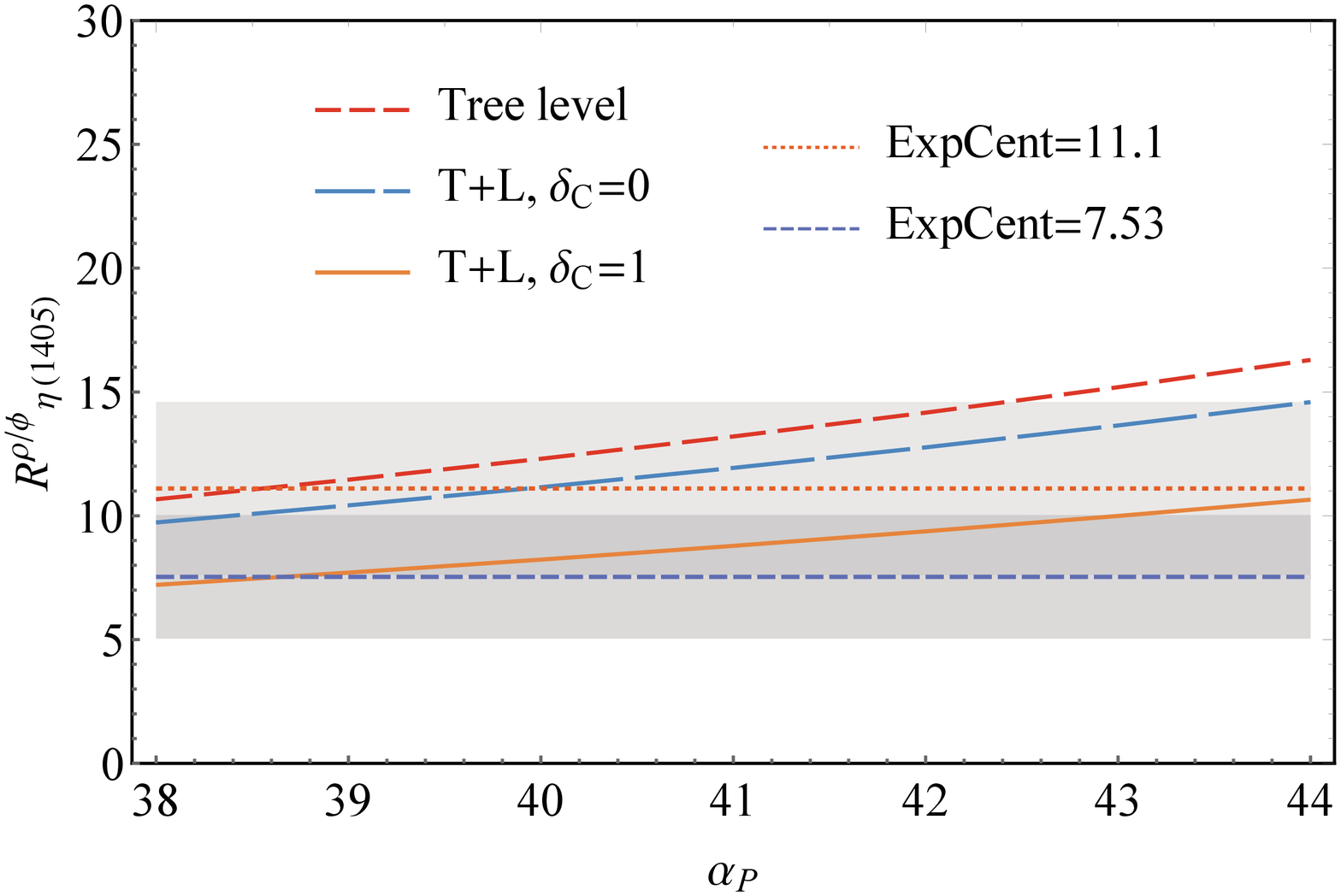}}
    \subfigure[]{\includegraphics[width=2.5 in]{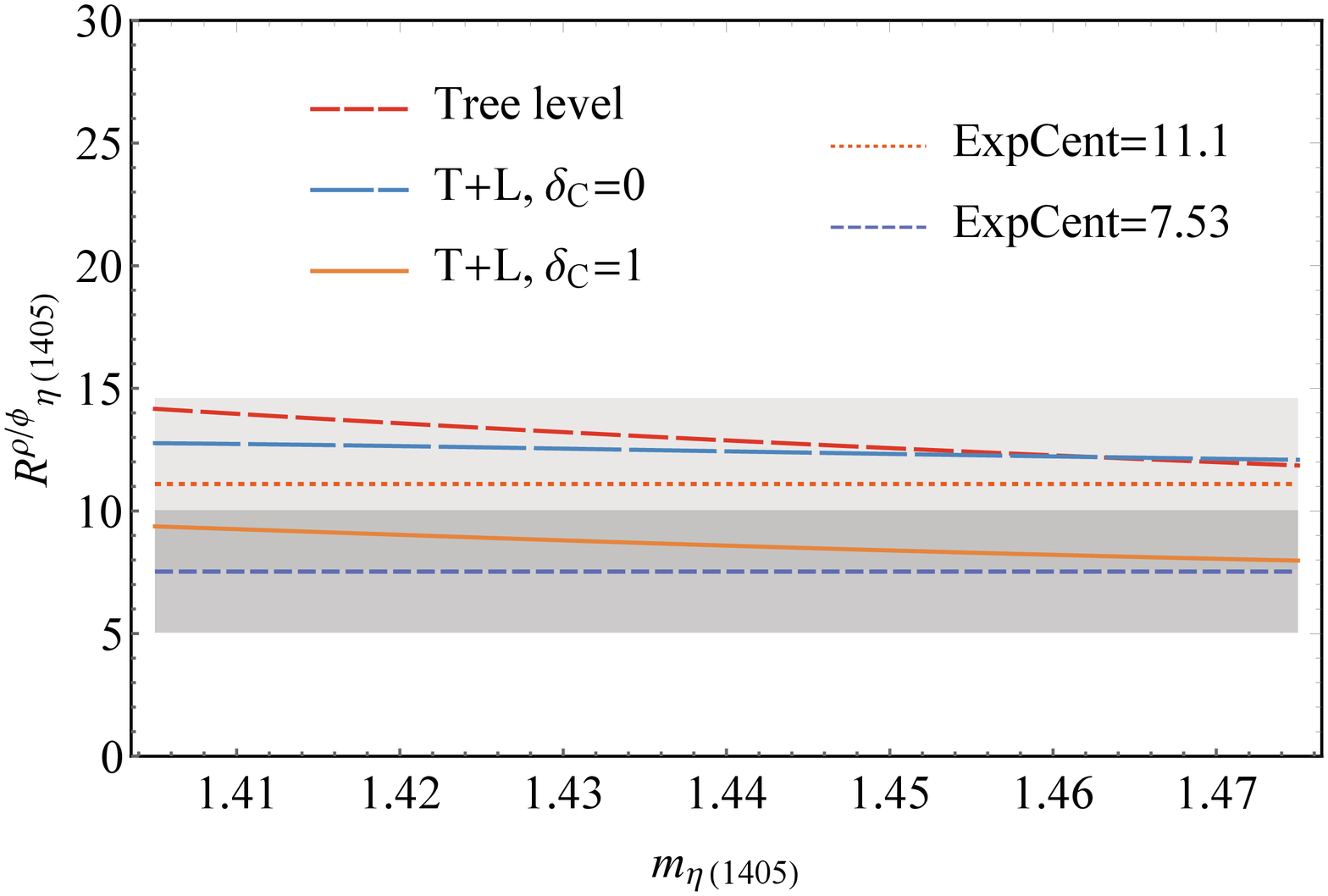}} 
    \caption{  (color online) (a) Dependence of $\mathcal{R}^{\rho/\phi}_{\eta(1405)}$ on the mixing angle $\alpha_P$ with $m_{\eta(1405)}$ fixed at $1.405$ GeV.   (b) Dependence of $ \mathcal{R}^{\rho/\phi}_{\eta(1405)}$ on the mass of $\eta(1405)$ with the mixing angle $\alpha_P$ set at $42^\circ$. In both figures the cut-off parameter $\alpha=1$ is adopted; ``T+L'' denotes the full calculations including the tree and loop contributions. $\delta_C$ is the coefficient of the contact diagrams and the values of $\delta_C=1$ and 0 will modulate the contributions from the contact loop diagrams. ``ExpCent'' denotes the central value of the two solutions from the BESIII analysis of $\mathcal{R}^{\rho/\phi}_{\eta(1405)}$, i.e. $11.10 \pm 3.50$ and $7.53 \pm 2.49$, respectively~\cite{Ablikim:2018hxj}. The light and dark grey bands indicate the error bars of the two solutions. }
  \label{fig-rho-phi-eta1405}
\end{figure}

In Fig.~\ref{fig-rho-phi-eta1405} the two overlapping bands denote the ranges of the experimental ratios extracted from the two solutions for $J/\psi\to \gamma \eta(1405/1475)\to \gamma\gamma\rho^0$ at BESIII~\cite{Ablikim:2018hxj} as listed in Tab.~\ref{table:experimentaldata}. The central values, $7.53 \pm 2.49$ and $11.10 \pm 3.50$, are denoted by the short-dashed and dotted lines, respectively. Within the commonly adopted values for $\alpha_P$, i.e. $\alpha_P\simeq 38^\circ\sim 44^\circ$ (Fig.~\ref{fig-rho-phi-eta1405} (a)), and within the mass region of $m_{\eta(1405)}=1.405\sim 1.475$ GeV (Fig.~\ref{fig-rho-phi-eta1405} (b)) the ratios (solid lines) are consistent with the data within the errors. We also plot the ratios with only the tree amplitude (middle-dashed lines) and tree plus triangle loop amplitudes (long-dashed lines) as a comparison. Again, the large discrepancies between the solid and long-dashed lines indicate that the loop diagrams involving the EM contact interactions (Fig.~\ref{fig-dynamics} (d)) provide the main contributions from the loop transitions.

\begin{figure}
  \centering
  \subfigure[]{\includegraphics[width=2.5 in]{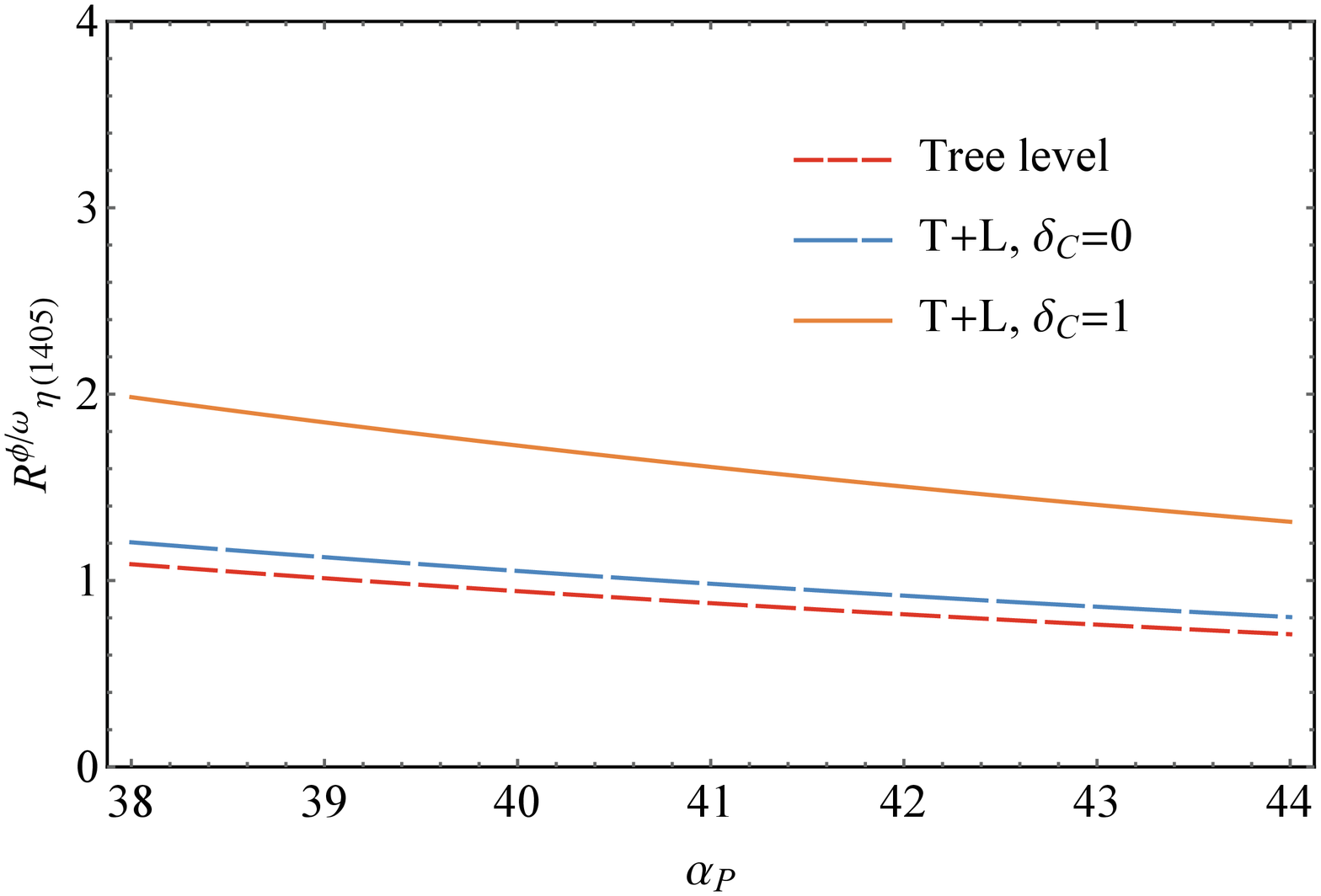}}
  \subfigure[]{\includegraphics[width=2.5 in]{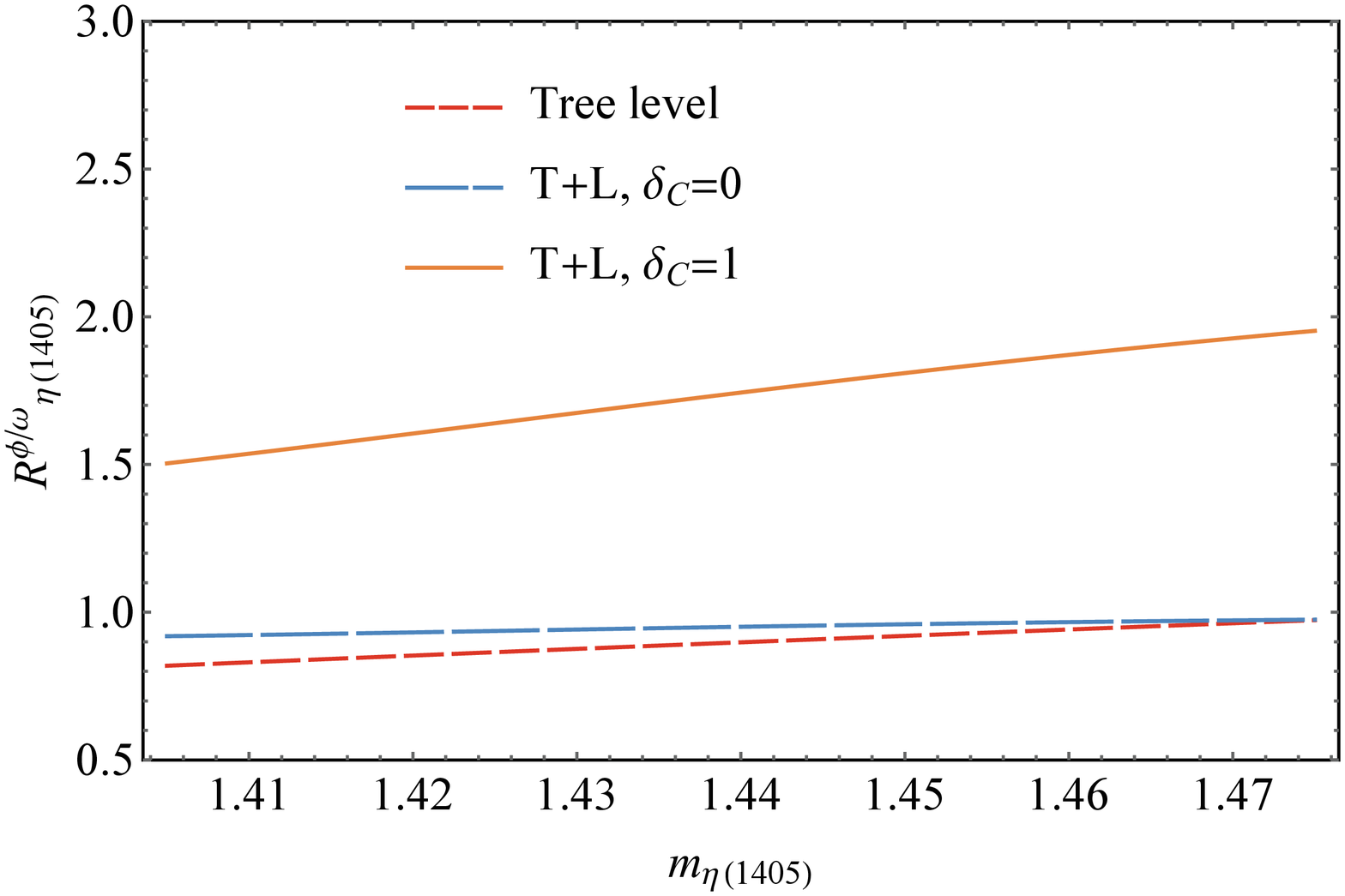}} 
  \caption{ (a) The dependence of  $ \mathcal{R}^{\phi/\omega}_{\eta(1405)}$ on the mixing angle $\alpha_P$ with $\alpha=1$ and $m_{\eta(1405)}=1.405$ GeV.  (b) The dependence of  $ \mathcal{R}^{\phi/\omega}_{\eta(1405)}$ on the mass of $\eta(1405)$ with the mixing angle $\alpha_P=42^{\circ}$. In both figures the cut-off parameter $\alpha=1$ is adopted. The legends of the curves are the same as those in Fig.~\ref{fig-rho-phi-eta1405}.}
  \label{fig-phi-omega-eta1405}
 \end{figure}

In Fig.~\ref{fig-phi-omega-eta1405} we present the predictions of $R^{\phi/\omega}_{\eta(1405)}$ for a range of $\alpha_P$ (Fig.~\ref{fig-phi-omega-eta1405} (a)) and $m_{\eta(1405)}$ (Fig.~\ref{fig-phi-omega-eta1405} (b)), respectively, similar to Fig.~\ref{fig-rho-phi-eta1405}. The ratios turn out to be stable and these two decays, i.e. $\eta(1405)\to\gamma \phi$ and $\gamma\omega$ are comparable with each other. Again, we see that the full amplitude calculations (solid lines) are significantly different from the results with only the tree diagram contributions  (short-dashed lines) or the tree plus the triangle loop amplitudes (long-dashed lines). This also indicates the dominant role played by the loop diagrams involving the EM contact interactions.

 \begin{figure}
  \centering
  \subfigure[]{\includegraphics[width=2.5 in]{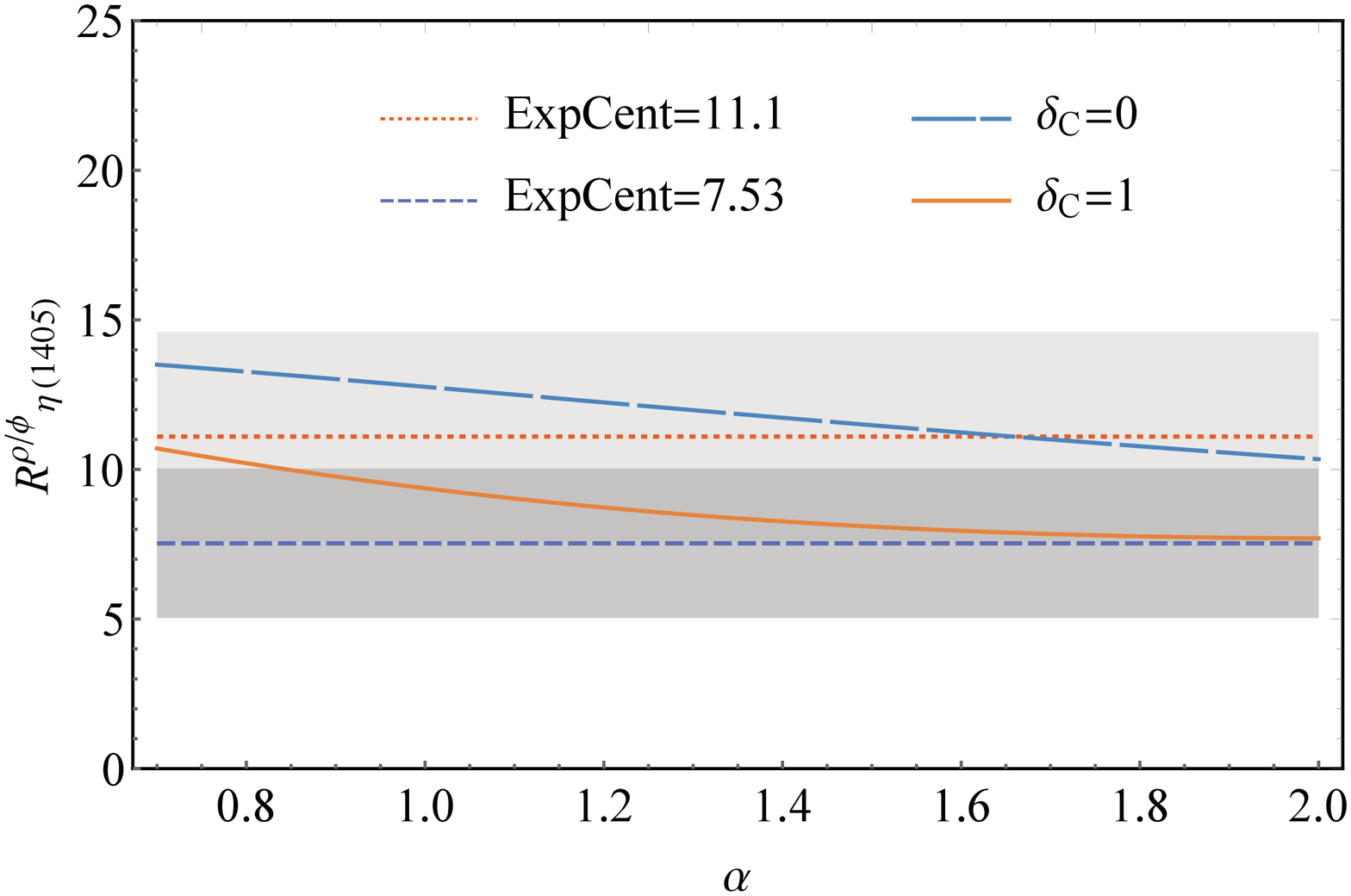}} 
  \subfigure[]{\includegraphics[width=2.5 in]{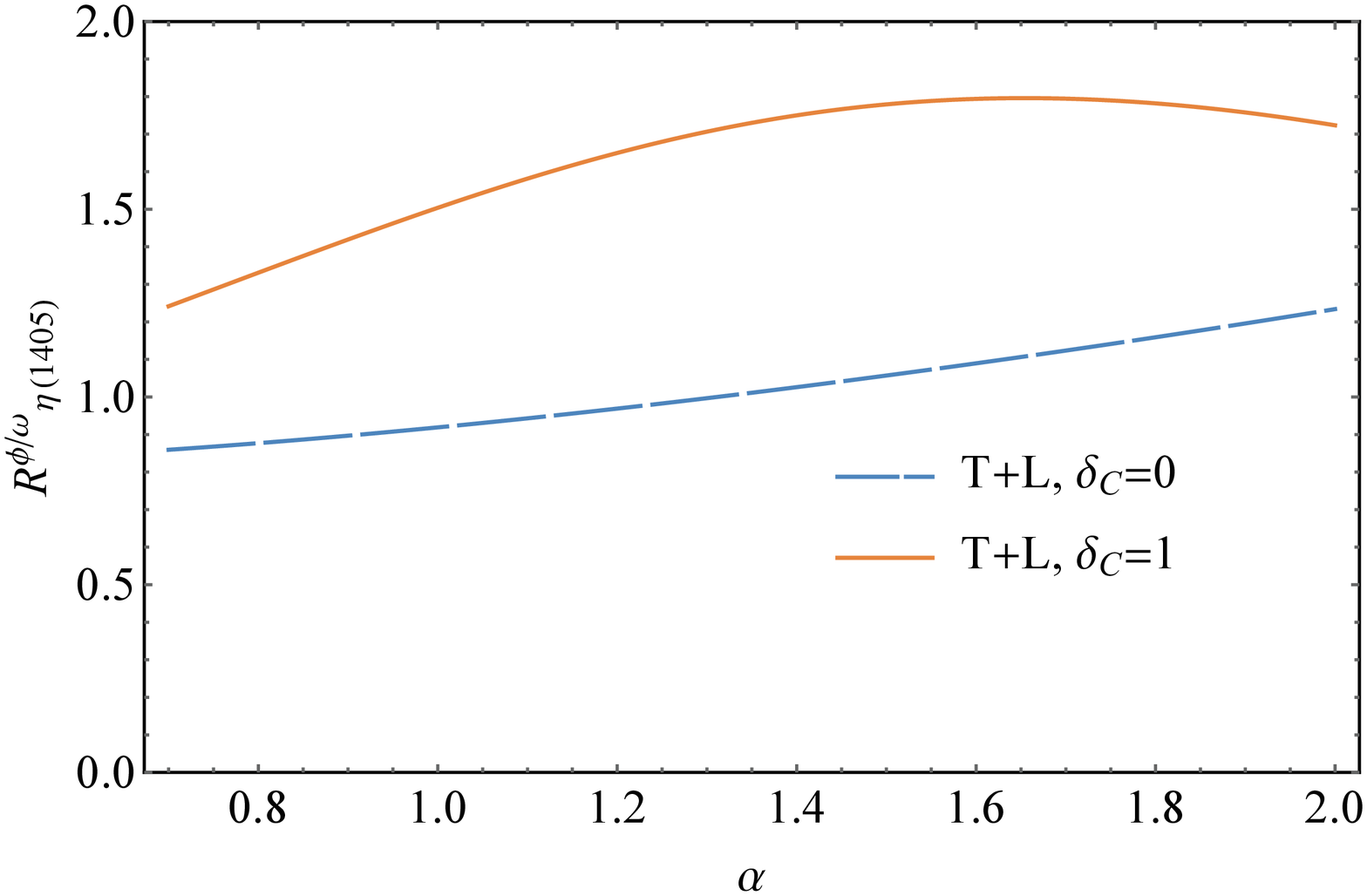}} 
  \caption{The dependence of $\mathcal{R} ^{\rho/\phi}_{\eta(1405)}$ (Fig. (a)) and $ \mathcal{R}^{\phi/\omega}_{\eta(1405)}$ (Fig. (b)) on the cut-off parameter $\alpha$, respectively, with $\alpha_P=42^{\circ}$ and $m_{\eta(1405)}=1.405$ GeV fixed in the calculations. The light and dark grey bands in (a) indicate the two solutions from the BESIII analysis, $7.53 \pm 2.49$ and $11.10 \pm 3.50$, respectively~\cite{Ablikim:2018hxj}.}
\label{fig-cutoff-eta1405}
\end{figure}

Another aspect to be examined is the cutoff dependence of the branching ratio fractions. In Fig.~\ref{fig-cutoff-eta1405}
both ratios $R^{\rho/\phi}_{\eta(1405)}$ and $R^{\phi/\omega}_{\eta(1405)}$ in terms of a range of the cut-off parameter $\alpha$ is presented in (a) and (b), respectively. It shows that $R^{\rho/\phi}_{\eta(1405)}$ is a stable quantity with the increasing $\alpha$, although it drops slightly within the range of $\alpha=1\sim 2$. In contrast, the ratio $R^{\phi/\omega}_{\eta(1405)}$ increase gradually in terms of $\alpha$. This indicates the increasing contributions of the loop amplitudes in $\eta(1405)\to\gamma\phi$ with the increasing $\alpha$. Combining Fig.~\ref{fig-cutoff-eta1405} (a) and (b) together, one can see that the relation $\Gamma_{\eta(1405)\to\gamma\rho^0}> \Gamma_{\eta(1405)\to\gamma\phi}> \Gamma_{\eta(1405)\to\gamma\omega}$ which is consistent with the expectation of the first radial excitation assignment~\cite{Wu:2011yx,Wu:2012pg}.

\subsubsection*{$\eta(1295)\to \gamma V$}

\begin{figure}
    \centering
    \subfigure[]{\includegraphics[width=2.5 in]{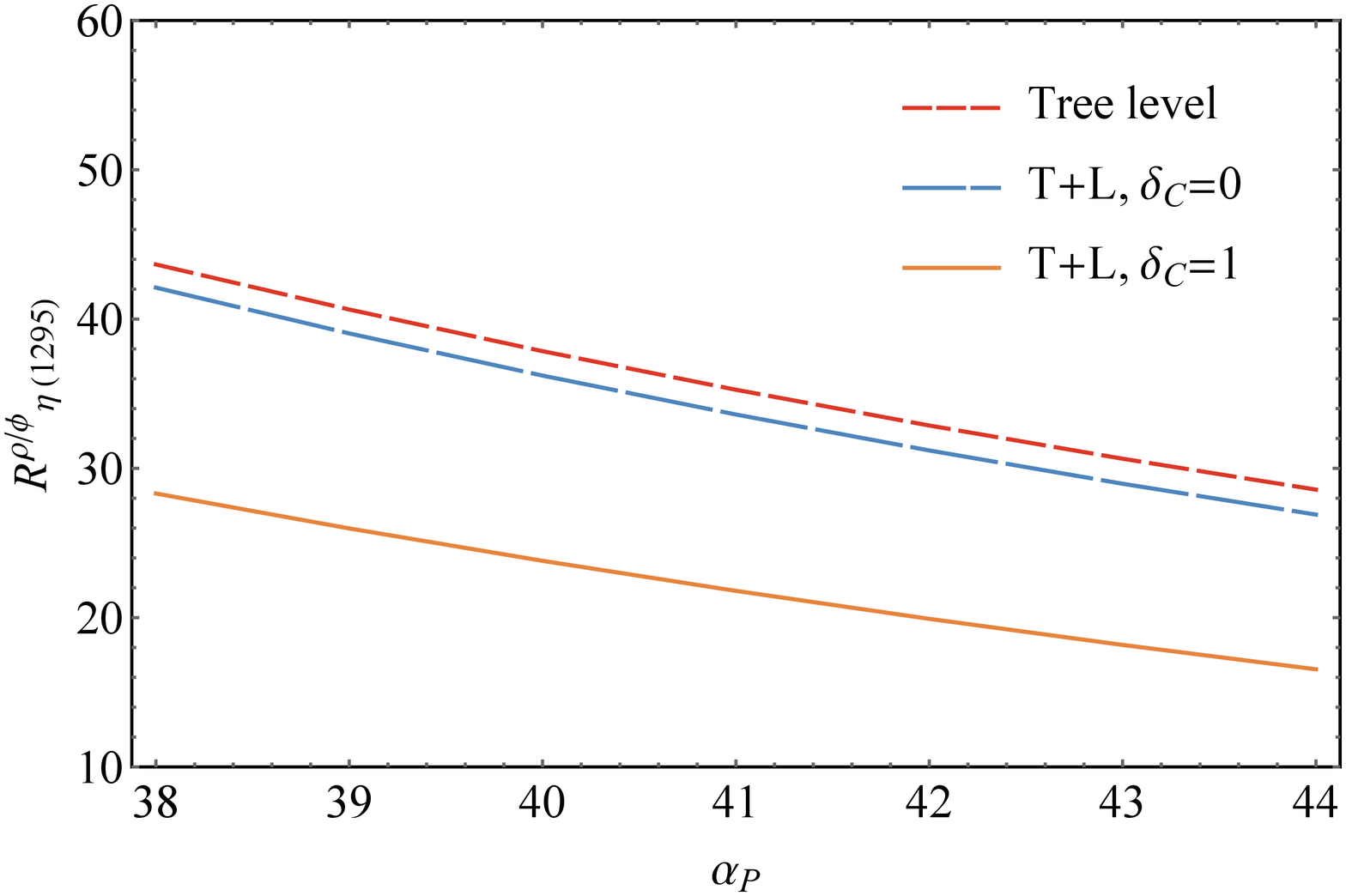}}  
    \subfigure[]{\includegraphics[width=2.5 in]{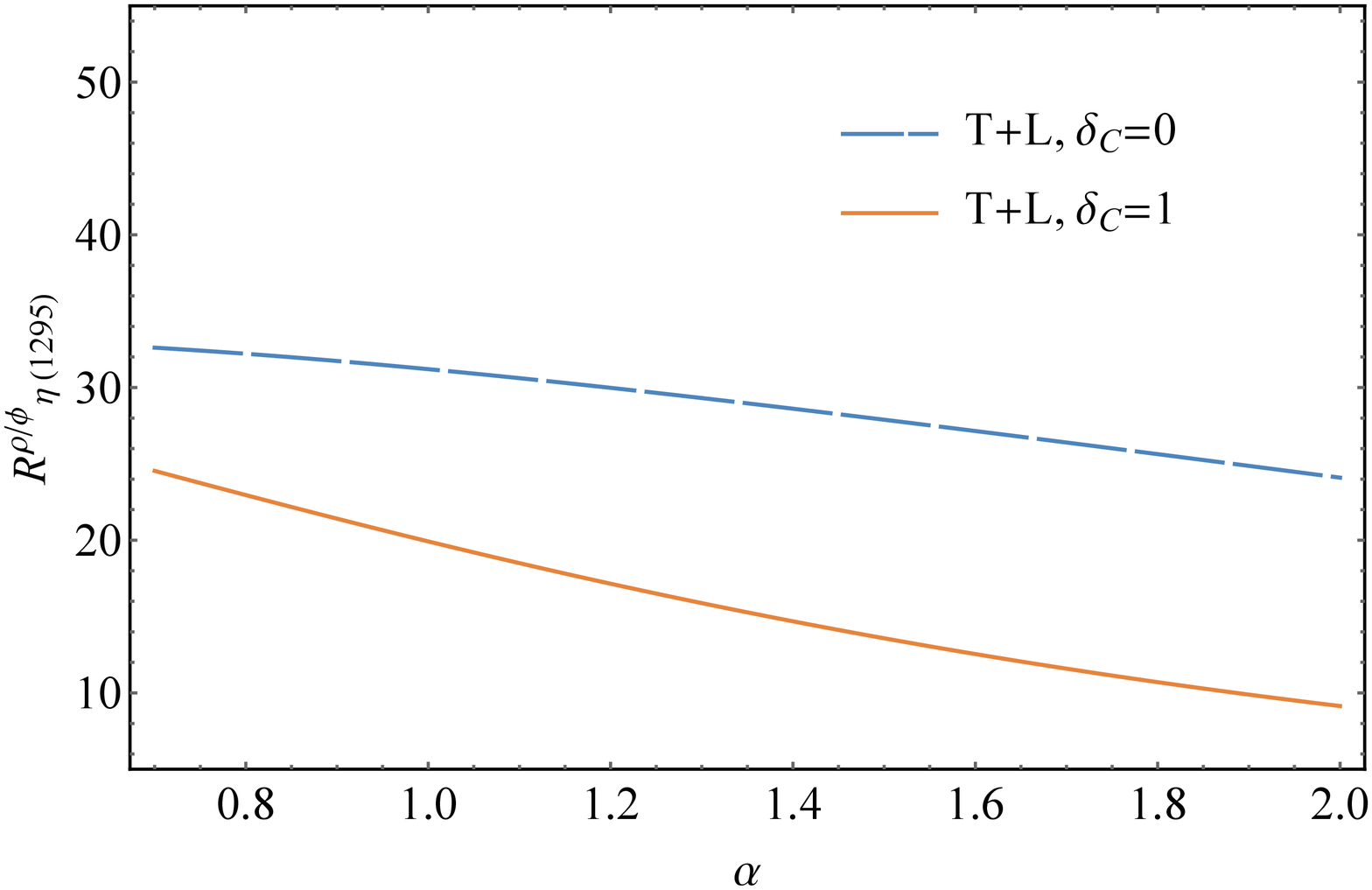}} 
     \caption{ (a)  The dependence of $ \mathcal{R}^{\rho/\phi}_{\eta(1295)}$ on the mixing angle $\alpha_P$ with $\alpha =1$. The middle dashed line is given by the tree-level transitions. (b)  The dependence of $\mathcal{R}^{\rho/\phi}_{\eta(1295)}$ on the cut-off parameter $\alpha$ with $\alpha_P =42^{\circ}$. In both calculations the solid and long-dashed lines denote the full calculations and calculations without the contact loop contributions, respectively.}
     \label{fig-rho-phi-1295}
  \end{figure}

In Fig.~\ref{fig-rho-phi-1295} (a) the predicted ratio $R^{\rho/\phi}_{\eta(1295)}$ in terms of the mixing angle $\alpha_P$ is plotted. The full calculation is denoted by the solid line while as a comparison, we also include the results for the exclusive tree-level transition (short-dashed line) and the tree-level amplitude plus the triangle transitions (long-dashed line). The difference between the long-dashed and solid lines indicates the significant contributions from the loop diagrams involving the EM contact interactions (see Fig.~\ref{fig-dynamics} (d)).

In Fig.~\ref{fig-rho-phi-1295} (b) the dependence of $R^{\rho/\phi}_{\eta(1295)}$ on the cut-off parameter $\alpha$ is shown where the solid and dashed line denote the full calculation and calculation without the the EM contact loop transitions, respectively. Within the range of $\alpha=1\sim 2$, the ratios keep to be stable although the ratios exhibit a decreasing tendency with the increasing $\alpha$.

\begin{figure}
      \centering
    \subfigure[]{\includegraphics[width=2.5 in]{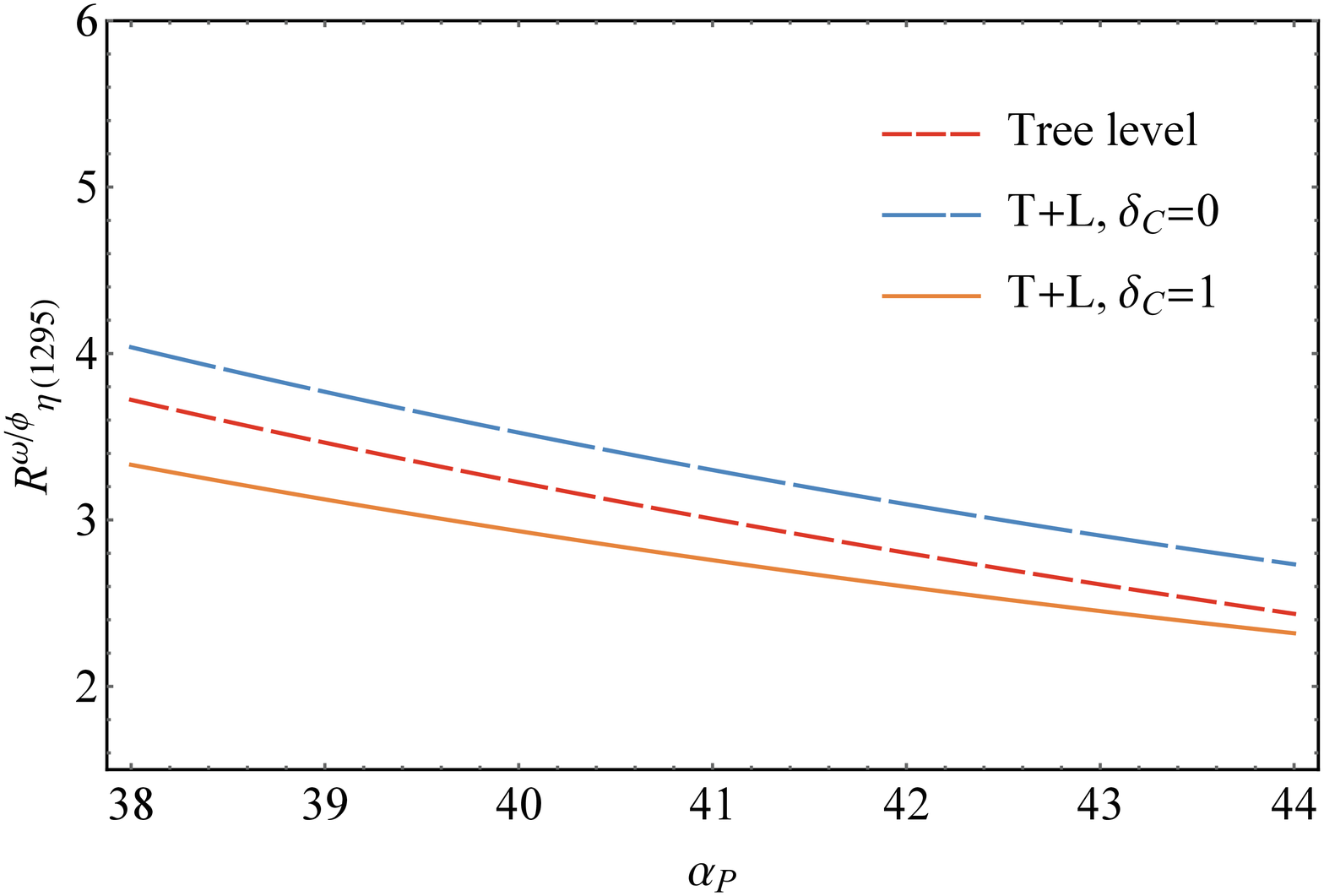}} 
     \subfigure[]{\includegraphics[width=2.5 in]{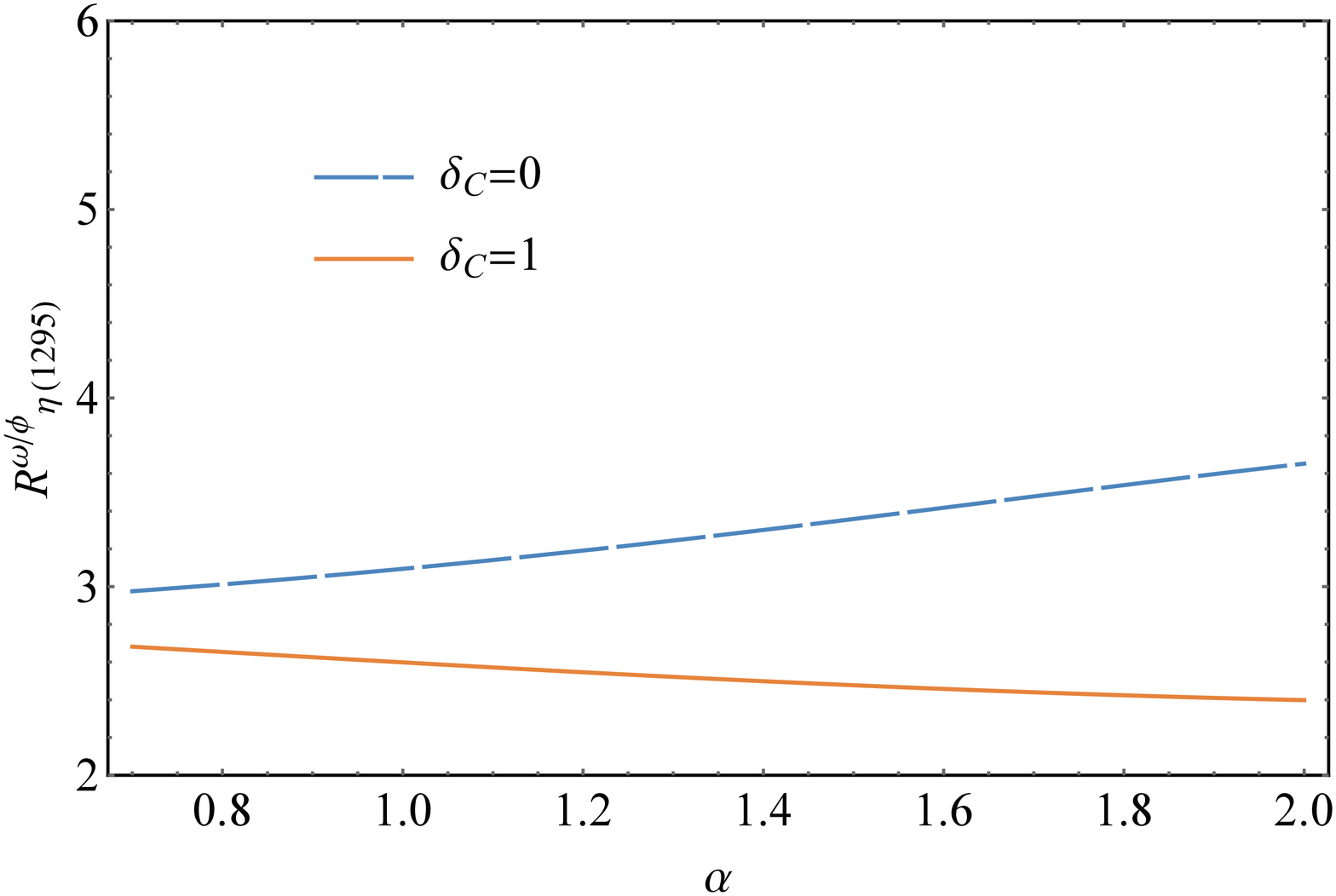}}  
     \caption{   (a) The dependence of $R^{\omega/\phi}_{\eta(1295)}$ on the mixing angle $\alpha_P$ with $\alpha=1$. 
    (b) The dependence of $R^{\omega/\phi}_{\eta(1295)}$ on parameter $\alpha$ with the mixing angle $\alpha_P=42^\circ$. The curve legends are the same as Fig.~\ref{fig-rho-phi-1295}.}
  \label{fig-omega-phi-1295}
  \end{figure}

In Fig.~\ref{fig-omega-phi-1295} the ratio $R^{\omega/\phi}_{\eta(1295)}$ is calculated and presented in the similar way as Fig.~\ref{fig-rho-phi-1295}. We can see that the dependence of the ratios on the mixing angle $\alpha_P$ (Fig.~\ref{fig-omega-phi-1295} (a)) is similar to  $R^{\rho/\phi}_{\eta(1295)}$. We also investigate the cut-off dependence of the loop transition contributions and the results are presented in Fig.~\ref{fig-omega-phi-1295} (b). It shows that the ratio without the EM contact loop contributions exhibits a smooth increase with the increasing $\alpha$. However, when the tree and loop amplitudes are properly included, one can see that the ratio keeps stable and insensitive to $\alpha$. 
      
\subsection{Loop influence on the mixing angle}

The above studies have shown the impact of the loop transitions on the experimental observables defined in the productions and decays of $\eta(1295)$ and $\eta(1405)$ in $J/\psi\to\gamma\gamma V$. As a consequence, it implies that the measured mixing angle between $\eta(1295)$ and $\eta(1405)$  may possess different values in different processes if the detailed transition mechanisms have not been properly included. 

To illustrate this, we adopt Eq.~(\ref{eq:rate-1405-rho-phi}) as an example. Without the loop transitions Eq.~(\ref{eq:rate-1405-rho-phi}) will define a mixing angle which can be extracted with the experimental value of $\mathcal{R}^{\rho/\phi}_{\eta(1405)}$ as the input. But if the loop transitions are included, the ratio should be expressed as 
\begin{eqnarray}\label{eq:rate-1405-rho-phi-mixing}
  \mathcal{R}^{\rho/\phi}_{\eta(1405)}&\equiv& \frac{\Gamma_{ \eta(1405)\to \gamma \rho}}{\Gamma_{\eta(1405)\to \gamma \phi}}=\left(\frac{|\mathbf{p}_\rho|}{|\mathbf{p}_\phi|}\right)^3\left[\frac{g^{T}_{\eta(1405) \gamma \rho}(\alpha_P)+g^{L}_{\eta(1405)\gamma \rho}(\alpha_P)}{g^{T}_{\eta(1405) \gamma \phi}(\alpha_P)+g^{L}_{\eta(1405)\gamma \phi}(\alpha_P)}\right]^2\frac{\tan^2\tilde{\alpha}_P}{2R^2} \ .
 \end{eqnarray}  
By matching it to the tree-level relation defined in  Eq.~(\ref{eq:rate-1405-rho-phi}), one has
\begin{eqnarray}
\left[\frac{(e m_\rho^2/f_\rho)G_\rho}{(e m_\phi^2/f_\phi)G_\phi}\right]^2\frac{\tan^2\tilde{\alpha}_P}{2R^2}&\equiv &
\left[\frac{g^{T}_{\eta(1405) \gamma \rho}(\alpha_P)+g^{L}_{\eta(1405)\gamma \rho}(\alpha_P)}{g^{T}_{\eta(1405) \gamma \phi}(\alpha_P)+g^{L}_{\eta(1405)\gamma \phi}(\alpha_P)}\right]^2 \ ,
\end{eqnarray}
and the ``empirical" mixing angle $\tilde{\alpha}_P$ can be extracted. In Table~\ref{tab-mixingangle} we list the extracted values of $\tilde{\alpha}_P$ from the ratios $R^{\rho/\phi}_{\eta(1405)}$ and $R^{\rho/ \phi}_{\eta(1295)}$. With three values for $\alpha_P$ in the calculations as an illustration, we see that the extracted values for the mixing angle $\tilde{\alpha}_P$ are different in the measurements of these two ratios. To some extent, the deviations of $\tilde{\alpha}_P$ from the commonly adopted values could be acceptable in a single channel. However, when put two channels together, such deviations should be regarded as significant. Further experimental measurements of these ratios would be able to clarify the role played by the loop transitions.

\begin{table}
  \centering
  \caption{  Extracting a new mixing angle $\tilde{\alpha}_P$ for the given mixing angle $\alpha_P$ with cut off parameter $\alpha=1$.}
  \begin{tabular}{c|c|c}
    \hline \hline
    \multirow{2}*{ $\alpha_P$ }   & \multicolumn{2}{c}{ $\tilde{\alpha}_P$ }                       \\ \cline{2-3}
                                               &   $R^{\rho/\phi}_{\eta(1295)}$   &   $R^{\rho/\phi}_{\eta(1405)}$  \\ \hline
$42^\circ$                           &          $49.2^\circ$                        &             $36.2^\circ$                         \\ \hline
$40^\circ$                           &          $46.7^\circ$                        &             $34.5^\circ$                         \\ \hline
$38^\circ$                           &          $44.1^\circ$                        &             $32.7^\circ$                         \\  
\hline \hline
  \end{tabular}
  \label{tab-mixingangle}
\end{table}

\section{Summary}\label{sec:4}

In this work, based on the one state assumption for $\eta(1405)$ and $\eta(1475)$, we systematically investigate the radiative decays of  $\eta(1295)$ and  $\eta(1405)$ by treating them as the first radial excitations of $\eta$ and $\eta'$. In the framework of the VMD model, we include the intermediate $\bar{K}K^*+c.c.$ meson loops as the leading correction to the tree-level transition amplitudes for the $\eta(1295)$ and  $\eta(1405)\to \gamma V$. With only one parameter, i.e. the cut-off parameter $\alpha$, we are able to to understand the production and decay behavior of both states in the $J/\psi$ radiative decays. In particular, the radiative decays of $\eta(1405)\to\gamma V$ can be described in agreement with the BESIII measurement. It is interesting to note that the loop transitions can produce significant effects in some decay channels. For instance, the loop contributions in $\eta(1295)\to \gamma\omega$ are found to be compatible with the tree-level contributions. This should not be surprising since the coupling of $g_{\eta(1295)K^*K}$ in the loop amplitude is sizeable and the $\omega$ meson decay constant in the tree-level amplitude is much smaller than that of the $\rho^0$ meson (see Tab.~\ref{tab-leptonicdecayconstant}). We also note that, due to the interference from the loop transition amplitudes, the production of $\eta(1295)$ in $J/\psi\to\gamma\gamma\rho^0$ will be significantly enhanced. Experimental study of $\eta(1295)$ in this channel is strongly recommended. 

In the scenario of assigning $\eta(1295)$ and  $\eta(1405)$ as the first radial excitation states of $\eta$ and $\eta'$, we show that the branching ratio fractions between these two states in the same decay channel, or between two exclusive decay channels for the same state, exhibit interesting patterns when the intermediate $\bar{K}K^*+c.c.$ meson loops are properly included. As a consequence of the loop corrections, if one tries to extract the mixing angle from the radiative decay data without taking into account the loop corrections, it may end up with values different from each other and different from the commonly adopted one for the $\eta$ and $\eta'$ mixing. We also find that the contributions from the meson loops are relatively small in $\eta(1405)\to\gamma\rho^0$. Thus, this channel will be dominated by the tree-level transition and is ideal for extracting the mixing angle. 

In brief, we find that the radiative decays of $J/\psi\to \gamma\gamma V$ can serve as a probe for understanding the nature of $\eta(1295)$ and $\eta(1405)$. With the large data sample of $J/\psi$ at BESIII, we can disentangle the role played by the meson loop transitions and gain more insights into the pseudoscalar meson spectrum.

\begin{acknowledgments}
This work is supported, in part, by the National Natural Science Foundation of China (Grant Nos. 11425525 and 11521505),  DFG and NSFC funds to the Sino-German CRC 110 ``Symmetries and the Emergence of Structure in QCD'' (NSFC Grant No. 12070131001, DFG Project-ID 196253076), National Key Basic Research Program of China under Contract No. 2020YFA0406300, and Strategic Priority Research Program of Chinese Academy of Sciences (Grant No. XDB34030302).
\end{acknowledgments}


\section*{Appendix: Amplitudes of the hadronic loop diagrams}
\begin{appendix}


In this Section, we present the loop amplitudes for the convenience of tracking the calculation details. 
For simplicity, we do not distinguish the coupling constants at the hadronic vertices, but just denote them as $g_i$ with $i=1, 2, 3$.

\begin{itemize}
  \item $[K^*,\bar{K},(K)]$ 
\end{itemize}

\begin{eqnarray}\label{appendixA_Ta}
  i \mathcal{M}
    &=& g_1 g_2 g_3 \int \frac{d^4p_1}{(2\pi)^4} (p_X +p_3)_\sigma \frac{(g^{\sigma\mu}-\frac{p_1^\sigma p_1^\mu}{p_1^2})}{p_1^2-m_{K^*}^2+i\epsilon}  \epsilon_{\alpha\beta\delta\mu} p_\gamma^\alpha p_1^\beta \epsilon_\gamma^\delta
  \frac{(p_2-p_3)_{\lambda}\epsilon_V^\lambda}{(p_2^2-m_K^2+i\epsilon)(p_3^2-m_K^2+i \epsilon)} \mathcal{F}(q_i^2)\nonumber \\
  &=& g_1 g_2 g_3 \int \frac{d^4 p_1}{(2\pi)^4 } \frac{\epsilon_{\alpha \beta \delta \mu} p^\alpha_\gamma p_1^\beta \epsilon^\delta_\gamma  (2 p_\gamma +2p_V -p_1)^{\mu} (2p_1-2p_\gamma -p_V)_\lambda \epsilon^\lambda_V}{(p_1^2-m_{K^*}^2+i\epsilon)(p_2^2-m_K^2+i\epsilon)(p_3^2-m_K^2+i \epsilon)} \mathcal{F}(q_i^2) \nonumber \\
  &=& 2  g_1 g_2 g_3 \epsilon_{\alpha\beta\delta\mu}p_\gamma^{\alpha} p_V^{\mu} \epsilon_\gamma^{\delta} \epsilon_V^{\lambda} \int \frac{d^4p_1}{(2\pi)^4} \frac{p_1^\beta[2(p_1)_\lambda-(2p_\gamma+p_V)_\lambda]}{(p_1^2-m_{K^*}^2+i\epsilon)(p_3^2-m_K^2+i\epsilon)(p_2^2-m_K^2+i \epsilon)} \mathcal{F}(q_i^2) \nonumber \\
  &=& 2  g_1 g_2 g_3 \epsilon_{\alpha\beta\delta\mu}p_\gamma^{\alpha} p_V^{\mu} \epsilon_\gamma^{\delta} \epsilon_V^{\lambda} \int \frac{d^4p_1}{(2\pi)^4} \frac{2p_1^\beta p_{1\lambda}}{(p_1^2-m_{K^*}^2+i\epsilon)(p_3^2-m_K^2+i\epsilon)(p_2^2-m_K^2+i \epsilon)} \mathcal{F}(q_i^2). 
\end{eqnarray}
Note that the due to the property of the antisymmetric tensor, only the $\delta^{\beta}_{\lambda }$ term could survive in the tensor integral. 
  
\begin{itemize}
  \item $[K^*,\bar{K},(K^*)]$ 
\end{itemize}

\begin{eqnarray}
i\mathcal{M} &=& \int \frac{d^4p_1}{(2\pi)^4} \frac{g_1 g_2 g_3(p_X+p_3)_\sigma(g^{\sigma \mu}-\frac{p^{\sigma}_1 p^{\mu}_1}{p^2_1})(g^{\rho\nu}-\frac{p^\rho_2 p^\nu_2}{p^2_2})\epsilon_{\alpha\beta\nu \lambda}p^\alpha_2p^\beta_V \epsilon^\lambda_V}
  {(p^2_1-m^2_{K^*})(p^2_2-m^2_{K^*})(p_3^2-m^2_{K})} 
    \times [\epsilon^{\delta}_{\gamma}g_{\delta\mu}p_{1\rho} +\epsilon^\delta_\gamma g_{\delta \rho}p_{2\mu} - g_{\mu\rho}(p_1+p_2)_\delta \epsilon^\delta_\gamma ] \mathcal{F}(p_i^2) \nonumber \\
    &=& g_1 g_2 g_3 \int  \frac{d^4p_1}{(2\pi)^4} \frac{-2}
    { p^2_1(p^2_1-m^2_{K^*})(p^2_2-m^2_{K^*})(p_3^2-m^2_{K})}   \nonumber \\ & & \times 
      \left\{  \epsilon_{\alpha \beta \nu \lambda} p_\gamma^\alpha p_V^\beta \epsilon_\gamma^\nu \epsilon_V^\lambda [(p_1 \cdot p_\gamma)^2-p_1^2(p_V \cdot p_\gamma)+(p_1 \cdot p_\gamma)(p_1\cdot p_V)]  \right.   \nonumber \\
      & & \left. -\epsilon_{\alpha \beta \nu \lambda} p_1^\alpha p_V^\beta \epsilon_\gamma^\nu \epsilon_V^\lambda [(p_1 \cdot p_\gamma)^2-p_1^2(p_V \cdot p_\gamma)+(p_1 \cdot p_\gamma)(p_1\cdot p_V)] \right. \nonumber \\
      & & - \left.  \epsilon_{\alpha \beta \nu \lambda} p_\gamma^\alpha p_V^\beta p_1^\nu \epsilon_V^\lambda [(p_1 \cdot \epsilon_\gamma -p_\gamma \cdot \epsilon_\gamma )(2p_1^2-p_1\cdot p_\gamma -p_1 \cdot p_V)-p_1^2(p_V \cdot \epsilon_\gamma)] \right\} \mathcal{F}(p_i^2).
 \end{eqnarray}
In the above integral, after contracting the Lorentz indices the amplitude can be simplified to a more compact form. One notices that the first term in the big bracket is a scalar integral, while the second and third term will pick up the linear terms containing $p_\gamma^\alpha$ and  $\epsilon_\gamma^\nu$, respectively, due to the property of the antisymmetric tensor. The same analysis is also applied to the following loop amplitudes.

 \begin{itemize}
  \item $[K^*,\bar{K}^*,(K)]$ 
\end{itemize}

\begin{eqnarray}
i \mathcal{M} &=& g_1 g_2 g_3 \int \frac{d^4 p_1}{(2\pi)^4}\frac{\epsilon_{\alpha\beta\mu\nu} p^\alpha_1 p^{\beta}_3 (g^{\mu \mu'}-\frac{p^{\mu}_1p^{\mu'}_1}{p^2_1})\epsilon_{\alpha_1\beta_1\mu'\delta}p^{\alpha_1}_1p^{\beta_1}_\gamma \epsilon^{\delta}_{\gamma}
  \epsilon_{\alpha_2 \beta_2 \nu' \lambda}p^{\alpha_2}_3 p^{\beta_2}_V \epsilon^{\lambda}_V (g^{\nu \nu'}-\frac{p^{\nu}_3p^{\nu'}_3}{p^{2}_3})}
  {(p^2_1-m^2_{K^*})(p^2_2-m^2_{K})(p^2_3-m^{2}_{K^*})} \mathcal{F}(q^2_i) \nonumber \\
  & = & g_1 g_2 g_3 \int \frac{d^4 p_1}{(2\pi)^4} \frac{\epsilon_{\alpha\beta\mu\nu}p^{\alpha}_1 p^{\beta}_3 \times \epsilon_{\alpha_1 \beta_1\mu \delta} p^{\alpha_1}_1 p^{\beta_1}_\gamma \epsilon^{\delta}_{\gamma}
  \times \epsilon_{\alpha_2 \beta_2 \nu \lambda} p^{\alpha_2}_3 p^{\beta_2}_V \epsilon^{\lambda}_V }{(p^2_1-m^2_{K^*})(p^2_2-m^2_{K})(p^2_3-m^{2}_{K^*})} \mathcal{F}(q^2_i) \nonumber \\
  & =& g_1 g_2 g_3  \int \frac{d^4 p_1}{(2\pi)^4} \frac{\mathcal{F}(q^2_i)}{(p^2_1-m^2_{K^*})(p^2_2-m^2_{K})(p^2_3-m^{2}_{K^*})} \times \left\{ \epsilon_{\alpha \beta \delta \lambda} p_\gamma^\alpha p_V^\beta \epsilon_\gamma^\delta \epsilon_V ^\lambda [(p_1 \cdot p_\gamma)^2+(p_1\cdot p_\gamma )(p_1\cdot p_\gamma)-p_1^2(p_V \cdot p_\gamma)] \right. \nonumber \\
  && +\epsilon_{\alpha \beta \delta \lambda} p_1^\alpha p_V^\beta \epsilon_\gamma ^\delta \epsilon_V^\lambda [-(p_1 \cdot p_\gamma)^2]+ \epsilon_{\alpha \beta \delta \lambda} p_\gamma^\alpha p_1^\beta \epsilon_\gamma ^\delta \epsilon_V^\lambda[(p_1 \cdot p_V)^2-p_1^2 p_V^2] \nonumber \\
  && -\epsilon_{\alpha \beta \delta \lambda} p_\gamma^\alpha p_V^\beta p_1 ^\delta \epsilon_V^\lambda[(p_1 \cdot p_\gamma)(p_1 \cdot \epsilon_\gamma)-(p_V \cdot p_\gamma)(p_1 \cdot \epsilon_\gamma)+(p_1 \cdot p_\gamma)(p_V\cdot \epsilon_\gamma)] \nonumber \\
  && \left. + \epsilon_{\alpha \beta \delta \lambda} p_\gamma^\alpha p_V^\beta \epsilon_\gamma ^\delta p_1^\lambda  [(p_1 \cdot p_V)(p_1 \cdot \epsilon_V)] \right \} 
\end{eqnarray}

\begin{itemize}
  \item $[K^*,\bar{K}^*,(K^*)]$ 
\end{itemize}

\begin{eqnarray}
  i \mathcal{M} &=& g_1 g_2 g_3 \int \frac{d^4 p_1}{(2\pi)^4}\frac{\epsilon_{\alpha\beta\mu\nu}p^\alpha_1 p^\beta_3 (g^{\mu{\mu'}}-\frac{p^\mu_1 p^{\mu'}_1}{p^2_1})
  (g^{\rho \sigma}-\frac{p^{\rho}_2 p^{\sigma}_2}{p^2_2})(g^{\nu{\nu'}}-\frac{p^{\nu}_3 p^{\nu'}_3}{p^2_3})}
  {(p^2_1-m^2_{K^*})(p^2_2-m^2_{K^*})(p^2_3-m^{2}_{K^*})} \nonumber \\
               &\quad&\times \bigg[ \epsilon^{\delta}_\gamma g_{\delta \mu'}p_{1\rho} +\epsilon^{\delta}_\gamma g_{\delta\rho} p_{2\mu'}-g_{\mu'\rho}(p_1+p_2)_{\delta} \epsilon^{\delta}_{\gamma}  \bigg]
 \times \bigg[ \epsilon^\lambda_{V} g_{\lambda\nu'}p_{3\sigma}-\epsilon^{\lambda}_V g_{\lambda \sigma} p_{2 \nu'}+g_{\nu'\sigma}(p_2-p_3)_\lambda \epsilon^\lambda_V\bigg] \nonumber \mathcal{F}(q^2_i), \\
 &=& g_1 g_2 g_3 \int \frac{d^4 p_1}{(2\pi)^4}\frac{\epsilon_{\alpha\beta\mu\nu}p^\alpha_1 p^\beta_3 
 (g^{\rho \sigma}-\frac{p^{\rho}_2 p^{\sigma}_2}{p^2_2})}
 {(p^2_1-m^2_{K^*})(p^2_2-m^2_{K^*})(p^2_3-m^{2}_{K^*})} \nonumber \\
              &\quad&\times \bigg[ \epsilon^{\delta}_\gamma g_{\delta \mu}p_{1\rho} +\epsilon^{\delta}_\gamma g_{\delta\rho} p_{2\mu}-g_{\mu\rho}(p_1+p_2)_{\delta} \epsilon^{\delta}_{\gamma}  \bigg]
\times \bigg[ \epsilon^\lambda_{V} g_{\lambda\nu}p_{3\sigma}-\epsilon^{\lambda}_V g_{\lambda \sigma} p_{2 \nu}+g_{\nu\sigma}(p_2-p_3)_\lambda \epsilon^\lambda_V\bigg] \nonumber \mathcal{F}(q^2_i), \\
   &=& g_1 g_2 g_3 \int \frac{d^4 p_1}{(2\pi)^4}  \frac{\mathcal{F}(q^2_i)}{p_2^2(p^2_1-m^2_{K^*})(p^2_2-m^2_{K^*})(p^2_3-m^{2}_{K^*})}  \nonumber \\
   &\quad&  \times \left\{ -\epsilon_{\alpha \beta \delta \lambda} p_1^\alpha p_V^\beta \epsilon_\gamma ^\delta \epsilon_V^\lambda[p_2^2(-p_1\cdot p_\gamma -p_1 \cdot p_V+p_1^2)+(p_1 \cdot p_\gamma-p_1^2)(-2(p_1 \cdot p_\gamma) +p_V \cdot p_\gamma -p_1 \cdot p_V +p_1^2)] \right. \nonumber \\
   &\quad&  +\epsilon_{\alpha \beta \delta \lambda} p_\gamma^\alpha p_1^\beta \epsilon_\gamma ^\delta \epsilon_V^\lambda [p_2^2(-p_1\cdot p_\gamma -p_1 \cdot p_V +p_1^2)+(p_1 \cdot p_\gamma-p_1^2)(-2(p_1 \cdot p_\gamma) +p_V \cdot p_\gamma -p_1 \cdot p_V +p_1^2)] \nonumber \\
   &\quad& -\epsilon_{\alpha \beta \delta \lambda} p_\gamma^\alpha p_V^\beta p_1 ^\delta \epsilon_V^\lambda[2 p_2^2(p_1\cdot \epsilon_\gamma)+p_2^2(p_1 \cdot \epsilon_\gamma -p_V \cdot \epsilon_\gamma)+(p_1 \cdot \epsilon_\gamma )(-2(p_1 \cdot p_\gamma)+p_V \cdot p_\gamma -p_1\cdot p_V+p_1^2)] \nonumber \\
   & &   + \epsilon_{\alpha \beta \delta \lambda} p_\gamma^\alpha p_V^\beta \epsilon_\gamma^\delta p_1^\lambda  [2 p_2^2(p_\gamma \cdot \epsilon_V)-3 p_2^2(p_1 \cdot \epsilon_V)-2(p_1^2-p_1\cdot p_\gamma)(p_1 \cdot \epsilon_V) \nonumber \\
   & &  \left.+ (p_1^2-p_1 \cdot p_\gamma)(p_1 \cdot \epsilon_V -p_\gamma \cdot \epsilon_V) +2(p_1^2-p_1\cdot p_\gamma)(p_\gamma \cdot \epsilon_V)] \right \} 
  \end{eqnarray}

\begin{itemize}
  \item $[K,\bar{K}^*,(K)]$ 
\end{itemize}

\begin{eqnarray}
  i \mathcal{M} &=& \int \frac{d^4 p_1}{(2\pi)^4} g_1 g_2 g_3 \frac{(p_X+p_1)_\mu (p_1+p_2)_\delta \epsilon^{\delta}_\gamma \epsilon_{\alpha\beta\nu\lambda}p^{\alpha}_3 p^{\beta}_V \epsilon^\lambda_V(g^{\mu \nu}-\frac{p^{\mu}_3 p^{\nu}_3}{p^2_3})}
  {(p^2_1-m^2_{K})(p^2_2-m^2_{K})(p^2_3-m^{2}_{K^*})} \mathcal{F}(q^2_i)\nonumber  \\
                       &=&  g_1 g_2 g_3\int \frac{d^4 p_1}{(2\pi)^4} \frac{\epsilon_{\alpha\beta\mu\lambda}p^{\alpha}_3 p^{\beta}_V \epsilon^{\lambda}_V(2p_1+p_3)^{\mu} (2p_1-p_\gamma)_\delta \epsilon^{\delta}_{\gamma} }
    {(p^2_1-m^2_{K})(p^2_2-m^2_{K})(p^2_3-m^{2}_{K^*})} \mathcal{F}(q^2_i) \nonumber \\
                       &=& g_1 g_2 g_3\int \frac{d^4 p_1}{(2\pi)^4} \frac{\epsilon_{\alpha\beta\mu\lambda}p^{\alpha}_3 p^{\beta}_V \epsilon^{\lambda}_V(2p_1)^{\mu} (2p_1)_\delta \epsilon^{\delta}_{\gamma} }
    {(p^2_1-m^2_{K})(p^2_2-m^2_{K})(p^2_3-m^{2}_{K^*})} \mathcal{F}(q^2_i) \nonumber \\
                       &=& g_1 g_2 g_3\int \frac{d^4 p_1}{(2\pi)^4} \frac{\epsilon_{\alpha\beta\mu\lambda}(p_\gamma+p_V-p_1)^\alpha p^{\beta}_V \epsilon^{\lambda}_V(2p_1)^{\mu} (2p_1)_\delta \epsilon^{\delta}_{\gamma} }
    {(p^2_1-m^2_{K})(p^2_2-m^2_{K})(p^2_3-m^{2}_{K^*})} \mathcal{F}(q^2_i) \nonumber\\
                       &=& 4 g_1 g_2 g_3  \epsilon_{\alpha\beta\mu\lambda} p^{\beta}_V \epsilon^{\lambda}_V \epsilon^{\delta}_{\gamma} \int \frac{d^4 p_1}{(2\pi)^4} \frac{p_\gamma^\alpha p_1^{\mu} p_{1\delta}-p_1^\alpha p_1^\mu p_1^\delta }
    {(p^2_1-m^2_{K})(p^2_2-m^2_{K})(p^2_3-m^{2}_{K^*})} \mathcal{F}(q^2_i)  \nonumber \\
    &=& 4 g_1 g_2 g_3  \epsilon_{\alpha\beta\mu\lambda} p_\gamma^\alpha p^{\beta}_V \epsilon^{\lambda}_V \epsilon^{\delta}_{\gamma} \int \frac{d^4 p_1}{(2\pi)^4} \frac{ p_1^{\mu} p_{1\delta}}
    {(p^2_1-m^2_{K})(p^2_2-m^2_{K})(p^2_3-m^{2}_{K^*})} \mathcal{F}(q^2_i)  \ .
\end{eqnarray}

\begin{itemize}
  \item $[K,\bar{K}^*,(K^*)]$ 
\end{itemize}

\begin{eqnarray}
 i \mathcal{M} &=& g_1 g_2 g_3 \int \frac{d^4 p_1}{(2\pi)^4} \frac{(p_X+p_1)_\mu \epsilon_{\alpha\beta\nu\delta}p^\alpha_2 p^{\beta}_\gamma \epsilon^{\delta}_{\gamma} 
    (g^{\nu \nu'}-\frac{p^{\nu}_2 p^{\nu'}_2}{p^2_2})  (g^{\mu'\mu}-\frac{p^{\mu'}_3 p^{\mu}_3}{p^2_3})}
    {(p^2_1-m^2_{K})(p^2_2-m^2_{K^*})(p^2_3-m^{2}_{K^*})}  \nonumber \\
    & &  \times [\epsilon^{\lambda}_V g_{\lambda \mu'}p_{3\nu'}-\epsilon^{\lambda}_V g_{\lambda\nu'}p_{2\mu'}+g_{\mu'\nu'}(p_2-p_3)_\lambda \epsilon^{\lambda}_V]  \mathcal{F}(q^2_i)  , \nonumber \\
    &=& g_1 g_2 g_3 \int \frac{d^4 p_1}{(2\pi)^4} \frac{(p_X+p_1)_\mu \epsilon_{\alpha\beta\nu\delta}p^\alpha_2 p^{\beta}_\gamma \epsilon^{\delta}_{\gamma} 
      (g^{\mu'\mu}-\frac{p^{\mu'}_3 p^{\mu}_3}{p^2_3})}
    {(p^2_1-m^2_{K})(p^2_2-m^2_{K^*})(p^2_3-m^{2}_{K^*})}  
 \times [\epsilon^{\lambda}_V g_{\lambda \mu'}p_{3\nu}-\epsilon^{\lambda}_V g_{\lambda\nu}p_{2\mu'}+g_{\mu'\nu}(p_2-p_3)_\lambda \epsilon^{\lambda}_V]  \mathcal{F}(q^2_i)  , \nonumber \\
   &=& g_1 g_2 g_3  \int \frac{d^4 p_1}{(2\pi)^4} \frac{ \mathcal{F}(q^2_i)}{p_3^2(p^2_1-m^2_{K})(p^2_2-m^2_{K^*})(p^2_3-m^{2}_{K^*})}   \nonumber \\
   & &  \times \left \{ -\epsilon_{\alpha \beta \delta \lambda} p_\gamma^\alpha p_1^\beta \epsilon_\gamma^\delta \epsilon_V^\lambda [2  (p_{\gamma }\cdot p_V)^2+2 p_1^2 \left( p_1\cdot p_{\gamma }\right)
   -p^2_3 \left(  p_{\gamma }\cdot p_V\right)+p^2_3 \left( p_1\cdot p_V\right) +p^2_3 p_1^2+p_1^2 \left(  p_{\gamma }\cdot p_V\right)-2 p_V^2 \left(p_1 \cdot p_{\gamma }\right) \right. \nonumber\\
   & &-2 \left(  p_1 \cdot p_V\right) \left( p_{\gamma } \cdot p_V\right)-4 \left( p_1 \cdot  p_{\gamma }\right) \left( p_{\gamma } \cdot p_V\right)+ p_V^2 \left(  p_{\gamma }\cdot p_V\right)
    +p_1^2 p_V^2+ p_1^2 \left( p_1 \cdot p_V\right)-p_V^2 \left( p_1 \cdot p_V\right)-p_1^4 ]  \nonumber \\
   & & - \epsilon_{\alpha \beta \delta \lambda} p_\gamma^\alpha p_V^\beta \epsilon_\gamma^\delta p_1^\lambda[-p^2_3 \left( p_{\gamma } \cdot \epsilon_V\right)+3 p^2_3 \left( p_1 \cdot \epsilon_V\right)-p_1^2 \left( p_{\gamma } \cdot \epsilon_V\right)-2 \left( p_{\gamma }\cdot p_V\right) \left( p_1 \cdot \epsilon_V\right)+ p_V^2 \left( p_{\gamma }\cdot  \epsilon_V\right)  \nonumber \\
   & & \left. +  2 \left( p_{\gamma } \cdot p_V\right) \left( p_{\gamma } \cdot \epsilon_V\right)+ p_1^2 \left( p_1 \cdot \epsilon _V\right)-p_V^2 \left( p_1 \cdot \epsilon _V\right) ]  \right\} .
\end{eqnarray} 

In the above amplitudes the product of the propagators and the form factor can be expanded as 
\begin{eqnarray}
  \frac{\mathcal{F}(q_i^2)}{ D_1 D_2 D_3} &=&\frac{1}{(p_1^2-m_1^2)(p_2^2-m_2^2)(p_3^3-m_3^3)}
  \bigg(\frac{m_1^2-\Lambda_1^2}{p_1^2-\Lambda_1^2}\bigg) \bigg(\frac{m_2^2-\Lambda_2^2}{p_2^2-\Lambda_2^2}\bigg) \bigg(\frac{m_3^2-\Lambda_3^2}{p_3^2-\Lambda_3^2} \bigg) \nonumber \\
   &= &\bigg(\frac{1}{p_1^2-m_1^2}-\frac{1}{p_1^2-\Lambda_1^2}\bigg)\bigg(\frac{1}{p_2^2-m_2^2}-\frac{1}{p_2^2-\Lambda_2^2}\bigg) \bigg(\frac{1}{p_3^2-m_3^2}-\frac{1}{p_3^2-\Lambda_3^2}\bigg) \nonumber \\
   &= &\frac{1}{(p_1^2-m_1^2)(p_2^2-m_2^2)(p_3^2-m_3^2)}-\frac{1}{(p_1^2-\Lambda_1^2)(p_2^2-m_2^2)(p_3^3-m_3^3)}-\frac{1}{(p_1^2-m_1^2)(p_2^2-\Lambda_2^2)(p_3^2-m_3^2)}  \nonumber \\
 & - &\frac{1}{(p_1^2-m_1^2)(p_2^2-m_2^2)(p_3^2-\Lambda_3^2)} 
  +\frac{1}{(p_1^2-\Lambda_1^2)(p_2^2-\Lambda_2^2)(p_3^2-m_3^2)}
  +\frac{1}{(p_1^2-\Lambda_1^2)(p_2^2-m_2^2)(p_3^2-\Lambda_3^2)}  \nonumber \\
  &+ &\frac{1}{(p_1^2-m_1^2)(p_2^2-\Lambda_2^2)(p_3^2-\Lambda_3^2)}-
  \frac{1}{(p_1^2-\Lambda_1^2)(p_2^2-\Lambda_2^2)(p_3^2-\Lambda_3^2)} \ . 
  \end{eqnarray}

\begin{itemize}
  \item $[K^*,\bar{K}]$ 
\end{itemize}
   
\begin{eqnarray}
i \mathcal{M} & =& e g_{\eta_X K^{*+}K^-} g_{V K^{*+}K^-}\int \frac{d^4 p_1}{(2\pi)^4} \frac{ (p_X+p_2)_\mu (g^{\mu\alpha}-\frac{p^{\mu}_1p^{\alpha}_1}{p^2_1})\epsilon_{\delta\beta\alpha\lambda}\epsilon^{\delta}_{\gamma}p^{\beta}_V \epsilon^{\lambda}_V}
    {(p^2_1-m^2_{K^*})(p^2_2-m^{2}_{K})} \mathcal{F}(q^2_i). \nonumber \\
    &=& e g_1 g_2 \int \frac{d^4 p_1}{(2\pi)^4} \frac{   \epsilon_{\delta\beta\alpha\lambda}\epsilon^{\delta}_{\gamma}p^{\beta}_V \epsilon^{\lambda}_V(p_X+p_2)_\mu (g^{\mu\alpha}-\frac{ p^{\mu}_1p^{\alpha}_1}{p^2_1}) }{(p^2_1-m^2_{K^*})(p^2_2-m^{2}_{K})} \mathcal{F}(q^2_i). \nonumber \\
    &=& e g_1 g_2 \int \frac{d^4 p_1}{(2\pi)^4} \left\{ \epsilon_{\alpha \beta \delta \lambda} p_1^\alpha p_V^\beta \epsilon_\gamma^\delta \epsilon_V^\lambda \frac{2(p_1\cdot p_\gamma+p_1 \cdot p_V)}{p_1^2 (p^2_1-m^2_{K^*})(p^2_2-m^{2}_{K}) }-\frac{2 \epsilon_{\alpha \beta \delta \lambda} p_\gamma^\alpha p_V^\beta \epsilon_\gamma^\delta \epsilon_V^\lambda}{(p^2_1-m^2_{K^*})(p^2_2-m^{2}_{K}) }  \right\} \mathcal{F}(q^2_i).
  \end{eqnarray}

\end{appendix}

\bibliographystyle{unsrt}

\end{document}